\documentstyle[preprint,eqsecnum,aps,epsf]{revtex}
\newcommand{\be}{\begin{equation}}
\newcommand{\ee}{\end{equation}}
\newcommand{\bea}{\begin{eqnarray}}
\newcommand{\eea}{\end{eqnarray}}
\newcommand{\bvec}[1]{\mbox{\boldmath $#1$}}

\begin{document}
\draft

\preprint{RCNP-Th/00041}

\title{Weyl symmetric representation of hadronic flux tubes\\
in the dual Ginzburg-Landau theory}

\author{Y. Koma$^{1,}$\footnote{Email address: koma@rcnp.osaka-u.ac.jp}, 
E. -M. Ilgenfritz$^{1}$, T. Suzuki$^{2}$, and H. Toki$^{1}$}

\address{
$^{1}$ Research Center for Nuclear Physics (RCNP), Osaka University,\\ 
Mihogaoka 10-1, Ibaraki, Osaka 567-0047, Japan\\
\vspace{0.3cm}
$^{2}$ Institute for Theoretical Physics, Kanazawa University, Kanazawa 920-1192, Japan}

\date{\today} 
\maketitle

\begin{abstract}\baselineskip = 0.6cm 
Hadronic flux-tube solutions describing the mesonic and
the baryonic states within the dual Ginzburg-Landau (DGL) theory
are investigated by using the dual lattice formulation 
in the Weyl-symmetric approach.
The shape of the flux tubes is determined by placement of 
the color-electric Dirac-string singularity treated 
as a connected stack of quantized plaquettes in the dual lattice formulation. 
The Weyl symmetric profiles of the hadronic flux tubes are obtained by 
using the manifestly Weyl invariant representation of the 
dual gauge field.\\ 
\end{abstract}


\section{Introduction}
\baselineskip = 0.6cm 

\par
The investigation of the dynamics of an Abrikosov vortex in the ordinary
superconductor is an important subject to understand superconductivity.
Now, we encounter quite a similar situation in the study of the
QCD vacuum and the hadron structure, since studies of 
lattice QCD in the maximally Abelian gauge 
\cite{Kronfeld:1987vd,Brandstater:1991sn,Bornyakov:1991es} show 
numerical evidence
of Abelian dominance \cite{Suzuki:1990gp,Bali:1996dm,Amemiya:1999zf}  
and monopole condensation 
\cite{Ivanenko:1993sc,Shiba:1995db,DelDebbio:1995yf,Chernodub:1997ps} for
the nonperturbative vacuum of QCD.
It means that the QCD vacuum can be considered as the dual superconductor 
\cite{Nambu:1974zg,Mandelstam:1974pi} described by the dual
Ginzburg-Landau (DGL) theory \cite{Suzuki:1988yq,Suganuma:1995ps}.
In this vacuum, the color-electric flux is squeezed into an almost
one dimensional object like a string due to the dual Meissner effect
caused by monopole condensation.
The various hadronic objects like the meson and the baryon are 
formed with the flux tubes existing in this picture. 
Hence it is very important to investigate the flux-tube dynamics 
in the QCD vacuum.

\par
The DGL theory can be obtained by using the Abelian 
projection \cite{'tHooft:1981ht}.
This scheme reduces the SU($N$) gauge theory to [U(1)]$^{N-1}$ Abelian 
gauge theory including color-magnetic monopoles. 
The symmetry [U(1)]$^{N-1}$ corresponds to the maximal 
torus subgroup of SU($N$).
Here, the dual gauge field is introduced by the Zwanziger 
formalism, which makes the electro-magnetic duality manifest 
in the presence of both electric charge and magnetic charge 
\cite{Zwanziger:1971hk}.
The summation over monopole world lines in four-dimensional space 
time can be rewritten as the theory of a complex scalar
field which interacts with the dual gauge field \cite{Bardakci:1978ph}.
Assuming monopole condensation, we finally get a Ginzburg-Landau type
Lagrangian with [U(1)]$^2$ dual gauge symmetry as an effective theory
of nonperturbative QCD 
\cite{Suzuki:1988yq,Suganuma:1995ps}.
According to the fact that QCD is a SU(3) gauge theory, 
there appear three different types of the Abelian color charges in 
the DGL theory, both in the electric sector and the magnetic sector, 
due to the Abelian projection. 
The color-electric charge and the color-magnetic charge
are defined in the weight vector and the root vector diagram, respectively, 
of the SU(3) algebra such as to satisfy the Dirac quantization condition.
These charges possess the global Weyl symmetry, which is 
permutation invariance among color labels of these charges, so that 
the Weyl invariance is an important aspect of the color-singlet criterion.

\par
In this paper, we study hadronic flux-tube solutions 
corresponding to mesonic and baryonic states in 
the [U(1)]$^2$ DGL theory, which are given by various kinds
of combination of the color-electric charges.
The baryonic state, which is composed of three
different types of the color-electric charge, is a characteristic new 
element of [U(1)]$^2$ DGL theory.
When we obtain the solution, we pay special attention to the 
Weyl symmetry, since the color-singlet state must be invariant under the 
Weyl transformation.
This can be achieved by using the manifestly Weyl symmetric representation 
of the dual gauge field.
We also study the usual Cartan representation with 3- and 8- 
basis for comparison \cite{Kamizawa:1993hb}.

\par
\begin{figure}[t]
\centerline{\epsfxsize=9.0cm\epsfbox{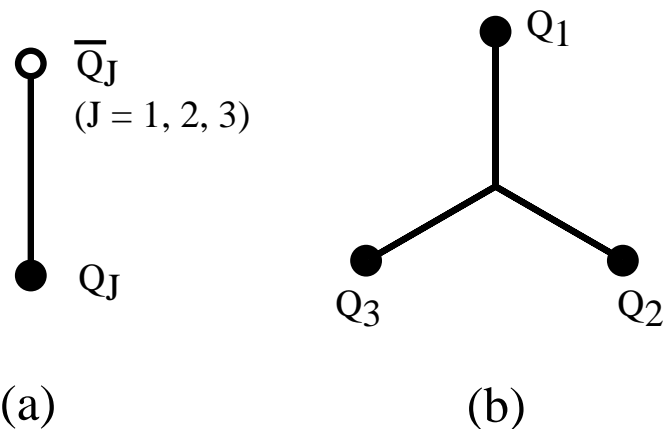}}
\caption{The possible combinations of the 
color-electric charge in the hadronic flux tubes, corresponding to 
(a) the meson and (b) the baryon.}
\label{fig:meso-bary-schematic}
\end{figure}

We first start from the U(1) DGL theory 
(dual Abelian Higgs model \cite{Nielsen:1973ve}) in order to 
get acquaintance with the general features of DGL theory.
The {\em dual lattice formulation} is introduced to obtain the 
various shapes of the flux-tube solution.
In the one-potential form of the DGL Lagrangian
in contrast to Zwanziger's two-potential form, 
there appears a nonlocal term which leads to the string-like 
singularity inside the flux tube. 
We call this color-electric Dirac string.
The lattice formulation is useful to treat the 
Dirac string singularity, since it is not 
defined on the dual lattice but on the ordinary lattice.
Next, we investigate the [U(1)]$^2$ DGL theory 
by using a similar but extended dual lattice formulation.
We discuss the various representations of the dual gauge field
including the Weyl symmetric representation.
Finally, we apply these formulation to systematically 
obtain the mesonic and the baryonic flux configurations 
[See, Fig.~\ref{fig:meso-bary-schematic}].

\section{The U(1) DGL theory (dual Abelian Higgs model) }

In order to warm up for the [U(1)]$^2$
DGL theory in the next section, we start from the U(1) DGL theory
with external quark sources, which is regarded to represent
the SU(2) gluodynamics in the Abelian projection. 
In this section, we mention some essential structures
of the dual lattice formulation for solving 
the non linear field equations for flux tubes.
In the presence of both electric and magnetic charges,
we have at least two forms of the Lagrangian, the Zwanziger 
form \cite{Zwanziger:1971hk} containing electric and magnetic 
vector potentials and the Blagojevic and Senjanovic (BS) 
form \cite{Blagojevic:1979bm}.
Although the Zwanziger form is useful to see the duality
between the electric sector and the magnetic sector, 
we adopt in this paper the BS form since in 
its one-potential form, written only in terms of the dual gauge field,
the flux-tube solutions are easy to see.

\subsection{The general feature}

\par
The U(1) DGL theory is given by the Lagrangian 
\footnote{In order to avoid confusion, `` $\hat{}$ '' is 
reserved for the parameters of U(1) DGL theory.}
\be
{\cal L}_{\rm U(1)\;DGL} =
-\frac{1}{4} {}^{*\!}F_{\mu \nu}^2 (B,j)
+ \left | \left ( \partial_{\mu} + i \hat{g} B_{\mu}\right )
\chi \right |^2  
-\hat{\lambda} \left ( |\chi|^2 - \hat{v}^2 \right )^2,
\label{eqn:DAHM-lagrangian}
\ee
where $B_{\mu}$ and $\chi$ are the dual gauge field and the complex
scalar monopole field, respectively.
The dual gauge coupling is $\hat{g}$, and $\hat{\lambda}$ 
characterizes the strength of monopole self interaction.
The monopole condensate $\hat{v}$ determines the mass scale of the system.
The dual field strength tensor ${}^{*\!}F_{\mu \nu}$ has the form
\be
{}^{*\!}F_{\mu \nu} (B,j)= 
\partial_{\mu} B_{\nu}-\partial_{\nu} B_{\mu}
-\frac{1}{n\cdot \partial} \epsilon_{\mu\nu\alpha\beta}
n^{\alpha}j^{\beta}.
\ee
Here, the nonlocal term appears as a contribution of the external 
quark current
\be
j_{\mu} = \frac{e}{2} \bar{q}\gamma_{\mu} q,
\label{eqn:u1-quarksrc}
\ee
where the factor $1/2$ is the weight of SU(2) algebra.
Accordingly, $e/2$ becomes the Abelian color-electric charge  
in SU(2) gluodynamics in the Abelian projection.
The nonlocal term is written more explicitly as
\be
\frac{1}{n \cdot \partial} \varepsilon_{\mu \nu \alpha \beta}
n^{\alpha} j^{\beta} (x)
=  
\int d^4 x' \langle x| \frac{1}{n \cdot \partial} |x' \rangle
\varepsilon_{\mu \nu \alpha \beta} n^{\alpha} j^{\beta}(x'),
\label{eqn:nonlocal-def}
\ee
where $\langle x| \frac{1}{n \cdot \partial} |x' \rangle $ 
is the kernel which satisfies the equation
\be
(n \cdot \partial)_x \langle x| \frac{1}{n \cdot \partial} |x' \rangle
= \delta^{(4)}(x-x').
\ee
Therefore, the solution is found to be
\bea
\langle x| \frac{1}{n \cdot \partial} |x' \rangle
&=&
\left [ p \theta ( (x- x')\cdot n )- (1-p) \theta ((x' - x)\cdot n )
\right ] \delta^{(3)} (\vec{x}_\perp - \vec{x}'_\perp ).
\label{eqn:string-kernel}
\eea
Here $p$ is an arbitrary real number and 
$\delta^{(3)}(x)$ is the $\delta$-function defined
on a three dimensional hyper-surface which has the normal vector $n_{\mu}$,
so that $\vec{x}_\perp$ and $\vec{x}'_\perp$ are three-vectors 
(generically not spatial) which are perpendicular to $n_{\mu}$.
One finds that this nonlocal term represents the string-like 
singularity, known as the {\em color-electric} Dirac string.
But now, there is only one type of color in U(1) DGL theory.
When we extend this idea to the [U(1)]$^2$ DGL theory,
we will have three types of color-electric Dirac strings.

\par
In the one-potential form of the U(1) DGL theory, the dual gauge 
field includes 
another color-electric Dirac string attached to the color-electric 
charge of a quark, which is canceled by the color-electric Dirac 
string in the nonlocal term.
In other words, the color-electric charge of the quark is defined by 
the cancellation of the color-electric Dirac string in the dual 
field tensor \cite{Ball:1988cf}.
Usually, such a singularity is considered to come from the 
phase of the monopole field, which is of course 
possible since they are related by singular dual gauge transformation. 
However, when we include the quark (color-electric charge) source 
in the theory, it seems natural to regard 
that the dual gauge field $B_{\mu}$ itself has a singular part
from the beginning.

\par
It should be noted that  the color-electric Dirac 
string is  `` dual '' to 
the original magnetic Dirac string which is 
attached to a magnetic monopole in the Abelian gauge theory like QED.
One may remember that the direction of a magnetic Dirac string 
can be varied by a singular Abelian gauge transformation, and hence,
the magnetic Dirac string is unphysical in the sense that 
one cannot detect it.
In our case, however, the symmetry which is responsible
for the direction of the color-electric Dirac string is 
the {\em dual gauge symmetry}, achieved by a set of transformation :
\bea
&&
\chi   \to \chi e^{if}, \quad
\chi^* \to \chi^* e^{-if},\quad
B_{\mu}\to B_{\mu}-\frac{1}{\hat{g}}\partial_{\mu}f,\nonumber\\
&&
-\frac{1}{\hat{g}}[\partial_{\mu},\partial_{\nu}] f
- \frac{1}{n \cdot \partial} \varepsilon_{\mu \nu \alpha \beta}
n^{\alpha} j^{\beta}
\to
\frac{1}{n' \cdot \partial} \varepsilon_{\mu \nu \alpha \beta}
n^{'\alpha} j^{\beta},
\label{eqn:u1-dualgaugesym}
\eea
where the dual gauge fixing function can be singular 
($[\partial_{\mu},\partial_{\nu}]f \ne 0$). 
The last relation in (\ref{eqn:u1-dualgaugesym}) determines the new
direction of the color-electric Dirac string $n_{\mu}'$.
This dual gauge symmetry would be broken by monopole condensation 
$\langle 0|\chi|0\rangle=\hat{v}$.
This is the so-called dual Higgs mechanism, which is realized by inserting  
$\chi = \left (\hat{v}+\phi/\sqrt{2} \right )e^{i\eta}\;$ 
(where $\phi,\eta \in \Re $) into the U(1) DGL Lagrangian as
\bea
{\cal L}_{\rm U(1)\;DGL} 
&=&
-\frac{1}{4} {}^{*\!}F_{\mu \nu}^2 (B',j)
+\frac{1}{2} m_B^2 B_{\mu}^{'2} 
+\frac{1}{2} \left [ (\partial_{\mu}\phi)^2 -m_\chi^2 \phi^2 \right ]
\nonumber\\
&&
+\hat{g}^2 B_{\mu}^{'2} \left ( \sqrt{2} \hat{v} \phi 
+ \frac{\phi^2}{2} \right )
-\hat{\lambda} \left (\sqrt{2}\hat{v} \phi^3 + \frac{\phi^4}{4} \right ),
\eea
where the phase of the monopole field $\eta$ is absorbed into the 
dual gauge field as $B'_{\mu} = B_{\mu} + \partial_{\mu}\eta/\hat{g}$,  
and accordingly,  the dual gauge field and the monopole field acquire 
the masses $m_B \equiv \sqrt{2}\hat{g}\hat{v}$ and 
$m_\chi \equiv 2\sqrt{\hat{\lambda}}\hat{v}$, respectively.
In that case, only the region where the field 
$\chi \approx 0$ resembles the normal phase in the 
dual superconductor vacuum, 
which means that the color-electric field can survive only 
near the region $\chi \approx 0$.
Then, the color-electric Dirac string has 
a physical meaning, since the `` normal region '' follows the color-electric 
Dirac string so as to minimize the energy of the system forming the 
color-electric flux tube.
It means that the shape and the size of this normal region
are determined by the direction $n_{\mu}$ and length, respectively.
The width of the flux tube is characterized 
by the inverse masses $m_B^{-1}$ and $m_\chi^{-1}$, 
which correspond to the penetration depth of the color-electric field 
and the coherence length of the monopole field, respectively.
The vacuum property, namely the type of dual superconductivity, is 
governed by the ratio of these lengths, 
the so-called Ginzburg-Landau (GL) parameter
\bea
\hat{\kappa} \equiv \frac{m_B^{-1}}{m_\chi^{-1}} 
= \frac{\sqrt{2 \hat{\lambda}}}{\hat{g}}.
\eea
Here, $\hat{\kappa}=1$ is the critical case, the so-called 
Bogomol'nyi limit, and the vacuum is classified into two 
types divided by this limit:
$\hat{\kappa} < 1$ belongs to the type-I vacuum and 
$\hat{\kappa} > 1$ is the type-II vacuum.
The profile functions connecting the normal phase in the center of 
the flux tube with the dual superconducting phase outside are  
classically determined by the field equations
\bea
&&
\partial^{\nu} {}^{* \!} F_{\mu \nu} =
-i \hat{g} 
\left ( 
\chi^{*}\partial_{\mu} \chi - \chi\partial_{\mu} \chi^{*}
\right )
+ 2 \hat{g}^2 B_{\mu} \chi^{*}\chi
\equiv k_{\mu},
\label{eqn:feq-u1-1}\\
&&
\left ( \partial_{\mu} +i \hat{g} B_{\mu} \right )^2 \chi
=
- 2 \hat{\lambda} \chi ( \chi^{*} \chi - \hat{v}^2),
\label{eqn:feq-u1-2}
\eea
where $k_{\mu}$ is the monopole {\em supercurrent} which circulates
in a transition region, confining the normal phase inside the dual 
superconducting phase.
Solving these field equations, the boundary condition are 
determined by the position of the color-electric Dirac string, 
and this information is included in the dual field strength tensor
as a nonlocal term.
In the following subsection, we explain the dual lattice formulation 
to solve such non linear equations and see how this formulation 
enables us to obtain various shapes of the flux-tube solutions.

\par
\subsection{The dual lattice formulation}

In order to formulate the U(1) DGL theory on the dual lattice,
taking into account (\ref{eqn:u1-quarksrc}), 
it is convenient to write the nonlocal term as
\be
\frac{1}{n \cdot \partial}\varepsilon_{\mu \nu \alpha \beta}
n^{\alpha} j^{\beta} \equiv \frac{e}{2} \Sigma_{\mu \nu},
\ee
where $\Sigma_{\mu \nu}$ denotes the singular field strength.
By using the relation (\ref{eqn:nonlocal-def}) and 
(\ref{eqn:string-kernel}), one finds 
that $\Sigma_{\mu \nu}$ satisfies
\be
\int_S \Sigma_{\mu \nu} d \sigma^{\mu\nu} = \pm 1,
\label{eqn:dirac-quantization}
\ee
where $S$ is a certain surface which is pierced by the color-electric 
Dirac string and the sign depends on the direction of the singular 
flux $\Sigma_{\mu \nu}$.  This is the Dirac quantization condition.
Accordingly, this also leads to the flux quantization condition of the 
color-electric flux induced by the dual gauge field.

\par
As long as we are interested in a three-dimensional static system, 
we can start from the Euclidean metric instead of the Minkowski metric.
Then, the U(1) DGL Lagrangian is written as
\be
{\cal L}_{\rm U(1) DGL}
= 
+\frac{1}{4} {}^{*\!}F_{\mu \nu}^2 (B,\Sigma)
+ \left | \left ( \partial_{\mu} + i \hat{g} B_{\mu}\right )
\chi \right |^2  + \hat{\lambda} 
\left ( |\chi|^2 - \hat{v}^2 \right )^2,
\ee
where 
\be
{}^{*\!}F_{\mu \nu} (B,\Sigma) 
= \partial_{\mu}B_{\nu} - \partial_{\nu}B_{\mu} 
-\frac{e}{2}\Sigma_{\mu\nu}.
\ee

\par
Now, we formulate this on the dual lattice. 
Let the dual gauge field be defined on {\em links}
as $B_{x,\mu}$,
the monopole field on {\em sites} as $\chi_x$, and 
the dual field strength tensor and the color-electric Dirac 
string term on {\em plaquettes} as ${}^{*\!}F_{x,\mu\nu}$ and 
$\Sigma_{x,\mu\nu}$.
The color-electric charge and anticharge are attached to the ends of the 
color-electric Dirac string.
We go over to dimensionless fields by the transformation : 
\bea
B_{x,\mu} \to \frac{\hat{B}_{x,\mu}}{a \hat{g}},\quad
\chi_x \to \hat{v}\hat{\chi}_x,\quad
\Sigma_{x,\mu\nu} \to \frac{\hat{\Sigma}_{x,\mu\nu}}{a^2},
\eea
where $a$ is dual lattice spacing, which has the dimension of 
length and can be reinserted when needed.
Accordingly, the scale is absorbed into the definition of
masses of the dual gauge field $m_B \equiv \sqrt{2}\hat{g} \hat{v}$, and
the monopole field $m_\chi \equiv 2\sqrt{\hat{\lambda}}\hat{v}$.
Then, the action on the dual lattice is given by
\be
S= \sum_{x} \hat{\beta}
\left [
\frac{1}{2} \sum_{\mu < \nu} {}^{*\!}\hat{F}_{x,\mu \nu}^2
+
\frac{\hat{m}_B^2}{2}
\sum_{\mu}
\left | \hat{\chi}_x - e^{i \hat{B}_{x,\mu}} 
\hat{\chi}_{x+\hat{\mu}} \right |^2
 +
\frac{\hat{m}_B^2 \hat{m}_\chi^2}{8}
\left ( |\hat{\chi}_x|^2 - 1 \right )^2
\right ],
\label{eqn:u1dgl-action}
\ee
where $\hat{\beta}\equiv 1/\hat{g}^2$, $\hat{m}_B \equiv m_B \cdot a$ 
and $\hat{m}_\chi \equiv m_\chi \cdot a$ \cite{Gubarev:1999yp}.
The dimensionless dual field strength with the external source 
is given by
\be
{}^{*\!} \hat{F}_{x,\mu \nu}
=
\hat{B}_{x,\mu} + \hat{B}_{x+\hat{\mu},\nu} 
-\hat{B}_{x+\hat{\nu},\mu} -  \hat{B}_{x,\nu}
- 2\pi \hat{\Sigma}_{x,\mu \nu},
\label{eqn:dual-field-tensor-u1}
\ee
where the relation between the color-electric charge and the 
color-magnetic charge, the Dirac quantization condition, $eg=4\pi$ is used.
The integral form of the color-electric flux quantization 
condition (\ref{eqn:dirac-quantization}) is then replaced by putting
\be
\hat{\Sigma}_{x,\mu \nu} = \pm 1,
\ee
on just a single plaquette in the $\mu \nu$ plane.
In the dual lattice formulation, the kinetic term of the monopole field
is written as
\bea
(\partial_{\mu} + i\hat{g}B_{\mu})\chi 
&\to&
\frac{\hat{v}}{a} \left ( U_{x,\mu} \hat{\chi}_{x+\hat{\mu}} 
- \hat{\chi}_x \right ),
\eea
where $U_{x,\mu}$ is a (compact) link variable,
\be
U_{x,\mu} = \exp \;(i a \hat{g} B_{x,\mu}) = \exp \;( i\hat{B}_{x,\mu}).
\ee
In the static three-dimensional system, we only need 
space-like links $\mu$ or $\nu =1, 2, 3$.
Note that four-dimensional Monte Carlo simulation of 
U(1) DGL theory in Euclidean metric is possible if we add 
the time-like link contribution.

\par
The field equation on the lattice is obtained 
when we formulate the cooling procedure,
which aims to minimize the action (\ref{eqn:u1dgl-action}). 
We require that the first derivative of the 
action  with respect to the dual gauge field and 
the monopole field becomes zero.
For the dual gauge field $\hat{B}_{x,i=1,2,3}$, this condition leads to 
\bea
\frac{\partial S}{\partial \hat{B}_{x,i}}
= \hat{\beta} 
\left ( {}^{*\!}\hat{F}_{x,ij} + {}^{*\!}\hat{F}_{x-\hat{j},ji} 
+ {}^{*\!}\hat{F}_{x,ik} + {}^{*\!}\hat{F}_{x-\hat{k},ki} 
+ m_B^2 b_{x,i}^{(2)} \right ) 
\equiv \hat{\beta} X_{x,i},\nonumber\\
\label{eqn:feq-u1-lat1}
\eea
which corresponds to  Eq.(\ref{eqn:feq-u1-1}) in the continuum limit.
Here we have defined
\bea
b_{x,i}^{(1)} &\equiv& \hat{\chi}_x^R 
\left ( 
 \hat{\chi}_{x+\hat{i}}^R \cos \hat{B}_{x,i}
-\hat{\chi}_{x+\hat{i}}^I \sin \hat{B}_{x,i}
\right )
+\hat{\chi}_x^I 
\left ( 
 \hat{\chi}_{x+\hat{i}}^R \sin \hat{B}_{x,i}
+\hat{\chi}_{x+\hat{i}}^I \cos \hat{B}_{x,i}
\right ),
\\
b_{x,i}^{(2)} &\equiv& \hat{\chi}_x^R 
\left ( 
 \hat{\chi}_{x+\hat{i}}^R \sin \hat{B}_{x,i}
+\hat{\chi}_{x+\hat{i}}^I \cos \hat{B}_{x,i}
\right )
-\hat{\chi}_x^I 
\left ( 
 \hat{\chi}_{x+\hat{i}}^R \cos \hat{B}_{x,i}
-\hat{\chi}_{x+\hat{i}}^I \sin \hat{B}_{x,i}
\right ).
\eea
The labels $i,j,k=1,2,3$ should be taken cyclically.
We find that the four terms of the dual field strength tensor 
${}^{*\!}\hat{F}_{x,ij} \sim {}^{*\!}\hat{F}_{x-\hat{k},ki}$ 
in (\ref{eqn:feq-u1-lat1}) are nothing else but the sum of 
plaquettes which are attached to the link at $x$ pointing into $i$-direction. 
The superscript of the monopole field $R,I$ denote its real and 
its imaginary part. 
The candidate value of the dual gauge potential, which locally 
satisfies the dual lattice field equations $X_{x,i}=0$,
is obtained by a relaxation step taking into account the second 
derivative of the action, {\it a la} Newton and Raphson as
\bea
\hat{B}_{x,i} \to
\hat{B}_{x,i}' &=& \hat{B}_{x,i} - 
\left ( \frac{\partial^2 S}{\partial \hat{B}_{x,i}^2} \right )^{-1}
\frac{\partial S}{\partial \hat{B}_{x,i} }\nonumber\\
&=&
\hat{B}_{x,i} - \frac{X_{x,i}}{4+ \hat{m}_B^2 b_{x,i}^{(1)}}.
\eea
For the monopole field, similarly, the local solution is given by
the update
\bea
\hat{\chi}_x^{R}
\to
\hat{\chi}_x^{R\;'}
&=&
\hat{\chi}_x^R -
\frac{X_x^R}{6 + \frac{1}{2} \hat{m}_\chi^2 
\left ( \hat{\chi}_x^{R\;2} +\hat{\chi}_x^{I\;2} -1 \right )} \nonumber\\
&&+
\frac{\hat{m}_\chi^2 \hat{\chi}_x^R 
\left ( \hat{\chi}_x^R X_x^R + \hat{\chi}_x^I X_x^I \right )}{
\biggl \{ 6 + \frac{1}{2} \hat{m}_\chi^2 
\left ( \hat{\chi}_x^{R\;2} +\hat{\chi}_x^{I\;2} -1 \right ) 
\biggr \}
\biggl \{ 6 + \frac{1}{2} \hat{m}_\chi^2 
\left ( \hat{\chi}_x^{R\;2} +\hat{\chi}_x^{I\;2} -1 \right ) 
+ \hat{m}_\chi^2 
\left ( \hat{\chi}_x^{R\;2} +\hat{\chi}_x^{I\;2}  \right )
\biggr \} },\nonumber\\
&&\nonumber\\
\hat{\chi}_x^{I}
\to
\hat{\chi}_x^{I\;'}
&=&
\hat{\chi}_x^I -
\frac{X_x^I}{6 + \frac{1}{2} \hat{m}_\chi^2 
\left ( \hat{\chi}_x^{R\;2} +\hat{\chi}_x^{I\;2} -1 \right )  } \\
&&+
\frac{\hat{m}_\chi^2 \hat{\chi}_x^I 
\left ( \hat{\chi}_x^R X_x^R + \hat{\chi}_x^I X_x^I \right )}{
\biggl \{ 6 + \frac{1}{2} \hat{m}_\chi^2 
\left ( \hat{\chi}_x^{R\;2} +\hat{\chi}_x^{I\;2} -1 \right ) 
\biggr \}
\biggl \{ 6 + \frac{1}{2} \hat{m}_\chi^2 
\left ( \hat{\chi}_x^{R\;2} +\hat{\chi}_x^{I\;2} -1 \right ) 
+ \hat{m}_\chi^2 \left (  \hat{\chi}_x^{R\;2} +\hat{\chi}_x^{I\;2} \right )
\biggr \}},\nonumber\\
\eea
where
\bea
X_x^R &=& 
6 \hat{\chi}_x^R 
-
\sum_{i=1}^3
\left \{
\left ( \hat{\chi}_{x+\hat{i}}^R \cos \hat{B}_{x,i} 
- \hat{\chi}_{x+\hat{i}}^I \sin  \hat{B}_{x,i}\right ) 
+
\left ( \hat{\chi}_{x-\hat{i}}^R \cos  \hat{B}_{x-\hat{i},i} + 
\hat{\chi}_{x-\hat{i}}^I \sin  \hat{B}_{x-\hat{i},i}\right ) 
\right \}\nonumber\\
&&
+
\frac{1}{2} \hat{m}_\chi^2 \hat{\chi}_x^R 
\left ( \hat{\chi}_x^{R\;2} +\hat{\chi}_x^{I\;2} -1 \right ), \\
&&\nonumber\\
X_x^I 
&=& 
6 \hat{\chi}_x^I 
-
\sum_{i=1}^3
\left \{
\left ( \hat{\chi}_{x+\hat{i}}^R \sin  \hat{B}_{x,i} 
+ \hat{\chi}_{x+\hat{i}}^I \cos  \hat{B}_{x,i}\right ) 
+
\left ( \hat{\chi}_{x-\hat{i}}^R (-\sin  \hat{B}_{x-\hat{i},i}) 
+ \hat{\chi}_{x-\hat{i}}^I \cos  \hat{B}_{x-\hat{i},i}\right ) 
\right \}
\nonumber\\
&&
+ \frac{1}{2} \hat{m}_\chi^2 \hat{\chi}_x^I 
\left ( \hat{\chi}_x^{R\;2} +\hat{\chi}_x^{I\;2} -1 \right ) .
\label{eqn:feq-u1-lat2}
\eea
The dual lattice field equation for the monopole field are $X_x^R=X_x^I=0$, 
which corresponds to Eq.~(\ref{eqn:feq-u1-2}) in the continuum limit.

\par
One finds that the behavior of the classical profile does not depend 
on the coupling $\hat{\beta}$, since this is factored out from the field 
equation.
Hence, one can set any $\hat{\beta}$ to study the behavior of profile.
At the same time, this implies that it is not necessary to specify
the lattice spacing $a$.
Once the masses $m_B$ and $m_\chi$ are provided in physical units, 
the lattice spacing $a$ is known to characterize
thickness and length of the flux tube.

\par
It is noted that when we discuss the magnitude of 
profiles or the classical string tension of the flux tube, 
$\hat{\beta}$ should be taken into account. 
In such case, $a$ also becomes important, 
since the dimensionful physical quantities are recovered by using this $a$.

\subsection{The solution}

\begin{figure}[t]
\centerline{\epsfxsize=10.0cm\epsfbox{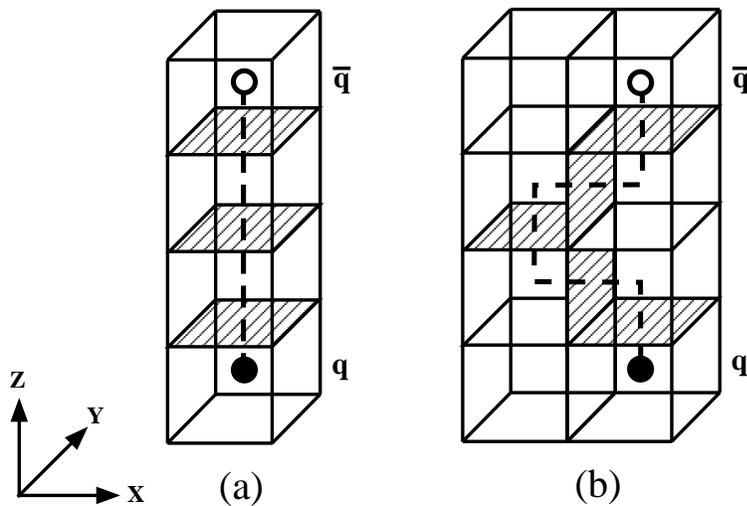}}
\caption{The color-electric Dirac string dual to singular 
plaquettes (shaded) ending in external charges.}
\label{fig:sigular-pla}
\end{figure}

Now, the boundary condition of the dual lattice field equations becomes 
very easy to handle, since all we have to do is to place a set 
of configurations of plaquettes $\hat{\Sigma}_{x,\mu\nu} \ne 0$
which is pierced by the color-electric Dirac string 
in the three dimensional space. 
For instance, if we assume that a straight color-electric Dirac string is 
placed on the $z$ axis, which means that the quark and the anti-quark
are placed on the $z$ axis, the only non-vanishing plaquette is 
$\hat{\Sigma}_{x,12}$, where $x=(0,0,x^3)$ and $x^3$ 
belongs to the region between a quark and an anti-quark.
A schematic figure is shown in Fig.~\ref{fig:sigular-pla}(a), where the 
non-vanishing plaquettes are shaded.
They form a connected stack of plaquettes dual to the color-electric 
Dirac string connecting $q$ and $\bar{q}$.
Here,  $\hat{\Sigma}_{x,\mu\nu} = +1$ $(-1)$ means that the color-electric Dirac 
string is regarded piercing the $\mu\nu$-plane at $x$ 
to $\mu \wedge \nu$ ($- \mu \wedge \nu$) direction.

\par
\begin{figure}[t]
\centerline{\epsfxsize = 14cm\epsfbox{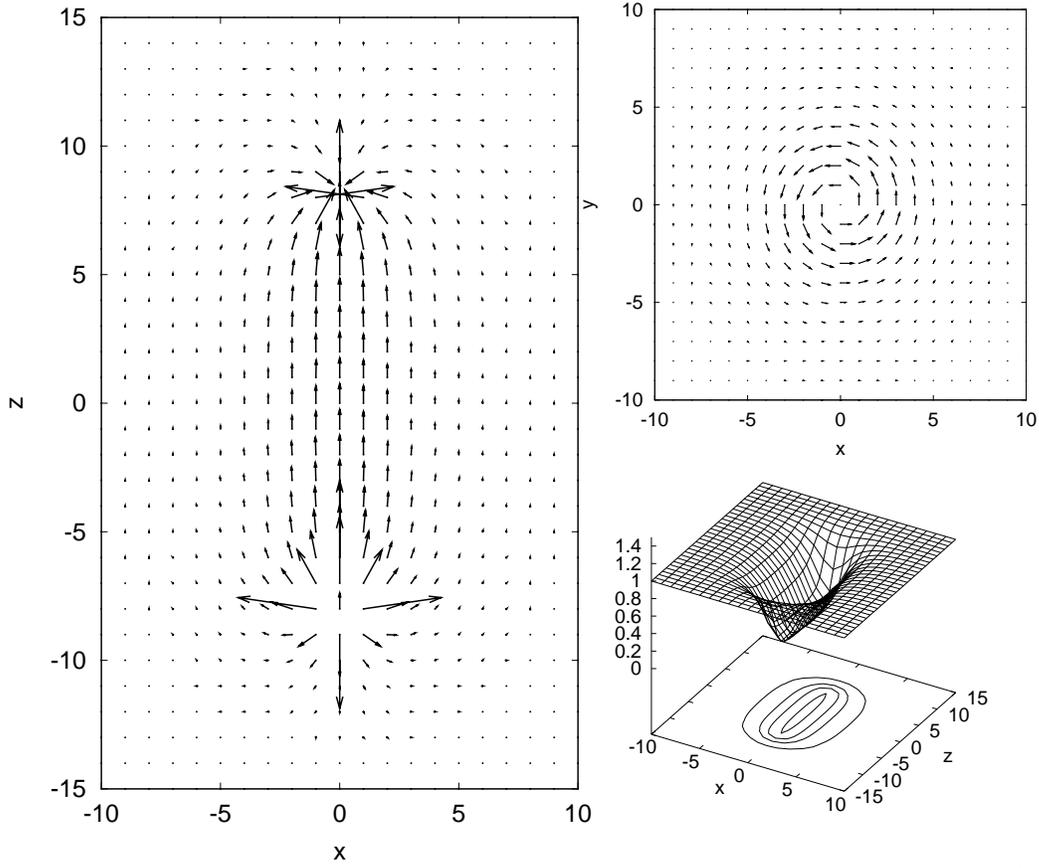}}
\caption{The profile of the color-electric field in the
$x$-$z$ plane at $y=0$ (left), 
the color-magnetic current in the $x$-$y$ plane at $z=0$ 
(right-upper), and the monopole field in the $x$-$z$ plane at $y=0$
(right-lower) of the mesonic flux tube in the U(1) DGL theory. 
The quark and the antiquark are placed at 
$(x,y,z)=(0,0,-8)$ and $(0,0,8)$, respectively.}
\label{fig:u1-meson-all}
\end{figure}

In Fig.~\ref{fig:u1-meson-all} we show the profiles of 
the color-electric field, 
the color-magnetic current which circulates around 
the flux tube, and the modulus of the monopole field.
Here a $32^3$ dual lattice is used, and the mass parameters are taken as
$\hat{\beta}=1$, $\hat{m}_B = \hat{m}_\chi = 0.5$.
The quark and the antiquark position are taken as
$(x,y,z) = (0,0,-8)$ and $(0,0,8)$, respectively.
The color-electric field is given by the space-space components of 
the dual field strength tensor (\ref{eqn:dual-field-tensor-u1}), 
${}^{*\!} \hat{F}_{x,ij}$ $(i,j=1,2,3)$.
The color-magnetic current is minus of the last term of 
$X_{x,i=1,2,3}$ in (\ref{eqn:feq-u1-lat1}), 
$-\hat{m}_B^2 b_{x,i}^{(2)}$, which corresponds to $k_{\mu}$ 
in (\ref{eqn:feq-u1-1}) in the continuum limit. 
The length of the arrows in these figures 
show the relative strength of fields.
In the figure of the color-electric field, one can observe
the Coulombic behaviors of the field at (near) the position of 
the quark (source) and the antiquark (sink).
Here, in order to obtain the vector variables defined on sites
from the color-electric fields 
on plaquettes and the color-magnetic currents on links, 
the appropriate averages like 
${}^{*\!}\hat{F}^{\rm fig}_{x,ij} \equiv  
({}^{*\!}\hat{F}_{x,ij}+{}^{*\!}\hat{F}_{x+\hat{k},ij})/2$,   
where $(i,j,k:cyclic)$, etc. are associated with centers of cubes. 
This is also where the quark and the antiquark are located.
Note that the parameter set used here is optimal for a $32^{3}$ dual lattice
and intended to compare with Ref.~\cite{Gubarev:1999yp}, 
where the relation of the flux-tube profile 
between the classical solution of U(1) DGL theory and that of 
the Abelian projected SU(2) lattice gauge theory \cite{Bali:1998gz} 
is discussed.

\par
\begin{figure}[t]
\centerline{\epsfxsize = 10cm\epsfbox{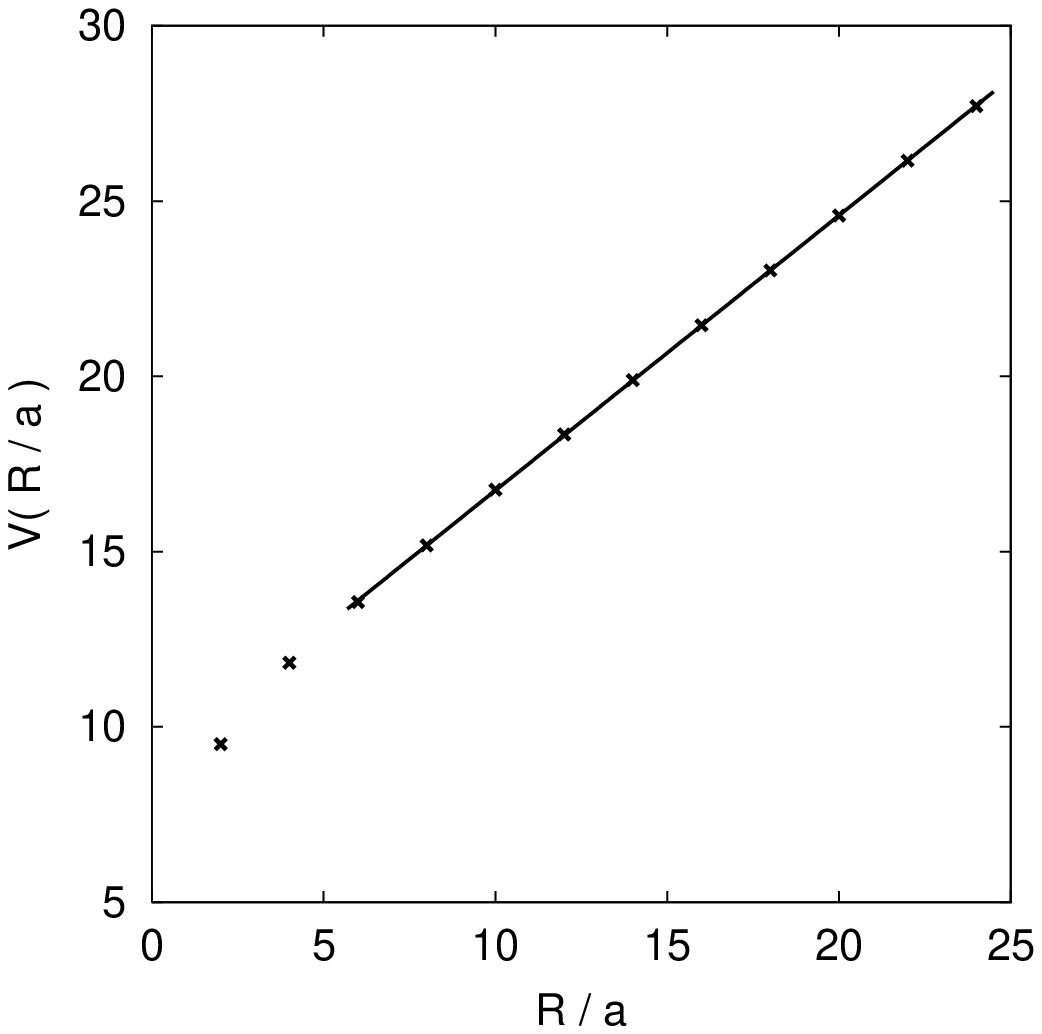}}
\caption{The quark-antiquark potential in the U(1) DGL theory, 
where $R/a$ denotes the $q$-$\bar{q}$ distance.
The parameter set is taken as $\hat{\beta} = 1$, 
$\hat{m}_B = \hat{m}_\chi =0.5$.}
\label{fig:u1-meson-pot}
\end{figure}

The relation $\hat{m}_B=\hat{m}_\chi$ implies that the vacuum is at 
the Bogomol'nyi limit, just between  type-I and type-II vacuum.
The inter-quark potential is shown in Fig.~\ref{fig:u1-meson-pot}. 
One finds that the slope of the linear part of the potential, 
which is the string tension, obeys the analytic result on the 
Bogomol'nyi limit, as $\sigma^L = 2 \pi \hat{v}^2 \cdot a^2 
= \hat{\beta} \pi \hat{m}_B^2 \sim  0.78$
\cite{Bogomolny:1976de,deVega:1976mi}. 
Here, the superscript ``~${}^L$~'' denotes the dimensionless string tension.
Note that the force always contains a Coulomb self-energy, 
which corresponds to a constant term in the potential $V(R/a)$.
If we choose a finer dual lattice, smaller $a$, the self 
energy becomes large, and accordingly, the constant takes a 
larger value. 
In such case, we could observe the fine structure of the 
short-distance behavior of the potential.
In this paper, we only pay attention to the long distance region.

\par
It is worth emphasizing that the dual lattice formulation 
presented here is also applicable to the `` bending '' 
flux tube [See Fig.~\ref{fig:sigular-pla}(b)].
If we assume that the bending is restricted in 1-3 plane, that means
that a $x^1$-component of the color-electric Dirac string
appears, {\it i.e.} some terms $\hat{\Sigma}_{x,23}$ have non-vanishing 
value, $\pm 1$.
In this case, the sign of this plaquette is similarly treated 
as discussed above.
In this sense, the dual lattice formulation is quite useful 
to obtain various shapes of the flux-tube solutions in U(1) DGL 
theory numerically.
In the next section, we investigate the [U(1)]$^2$ DGL theory
with the similar technique. In the [U(1)]$^2$ DGL theory,
there appears a flux-tube structure 
which includes three valence quarks corresponding to the baryonic state.
In order to study such a flux configuration, we need the skill to 
treat the bending flux tube.

\section{The [U(1)]$^2$ DGL theory}

In this section, we analyze the [U(1)]$^2$ DGL theory by using a similar 
technique as in  the previous section.
The main difference from the U(1) DGL theory is now that the 
dual gauge symmetry is extended to [U(1)]$^2$, which corresponds to  
Abelian projected SU(3) gluodynamics.
Accordingly, there appear three different types
of color-electric charge and color-magnetic charge, 
respectively.
Among these charges, we have the global {\em Weyl} symmetry,  
which is permutation invariance of the color charges.

\subsection{The general feature}

The [U(1)]$^2$ DGL Lagrangian in the one-potential form 
similar to the  U(1) case is given by 
\footnote{In the [U(1)]$^2$ DGL theory, we do not use `` $\hat{}$ '' 
for the parameters in order to distinguish from the U(1) ones.}
\bea
{\cal L}_{\rm [U(1)]^2 DGL} &=&
-\frac{1}{4}{}^{*\!} \vec{F}_{\mu\nu}^2(\vec{B},\vec{j})
+\sum_{i=1}^3
\left [ \left | \left (\partial_{\mu}+
ig\vec{\epsilon}_{i}{\cdot}\vec{B}_{\mu} \right )\chi_{i} \right |^2
-\lambda \left ( \left |\chi_{i} \right |^2-v^2 \right )^2 \right ],
\label{eqn:DGL}
\eea
where the dual field tensor has the form
\be
{}^{*\!} \vec{F}_{\mu\nu}(\vec{B},\vec{j})=
\partial_{\mu}\vec{B}_{\nu}-\partial_{\nu}\vec{B}_{\mu}
-\frac{1}{n \cdot \partial}\varepsilon_{\mu \nu \alpha \beta}
n^{\alpha} \vec{j}^{\beta}.
\ee
In this Lagrangian, 
$\vec{B}_{\mu}$ and $\chi_{i}$ denote the two-component
dual gauge field and the three-component complex scalar monopole field.
The interaction between quarks through the dual gauge field and with the 
monopole field originates from the existence of a quark current 
$\vec{j}_{\mu} = e \bar{q}\gamma_{\mu} \vec{H} q$ in the nonlocal term, 
where $\vec{H}=(T_{3},T_{8})$. 
Since the quark field is regarded as a fundamental representation of 
SU(3) group, this has a form
\be
q=\left(
\begin{array}{c}
q_1\\
q_2\\
q_3
\end{array}
\right),
\ee
where the labels 1, 2, 3 correspond to the three types of the 
color-electric charge red ($R$), blue ($B$) and green ($G$).
By using the relation
\be
\vec{H}=(T_3,T_8)=
\left(
\begin{array}{ccc}
\vec{w}_1 & 0         & 0\\
0         & \vec{w}_2 & 0\\
0         & 0         & \vec{w}_3 
\end{array}
\right),
\ee
where $\vec{w}_{j}$ are the weight vectors of the SU(3) algebra,
\be
\vec{w}_1= \left (\frac{1}{2}, \frac{1}{2\sqrt{3}} \right ),\quad 
\vec{w}_2= \left(-\frac{1}{2}, \frac{1}{2\sqrt{3}} \right ),\quad 
\vec{w}_3= \left (0, -\frac{1}{\sqrt{3}} \right ),
\ee
we obtain an explicit form of the quark current,
\be
\vec{j}_{\mu}=e \sum_{j=1}^3 \vec{w}_j \; \bar{q}_j \gamma_{\mu} q_j.
\label{eqn:cur-general}
\ee
We find that the color-electric charge is given by $e\vec{w}_j$.
The nonlocal term can be written in the similar way 
as U(1) DGL theory as 
\bea
\frac{1}{n \cdot \partial} \varepsilon_{\mu \nu \alpha \beta}
n^{\alpha} \vec{j}^{\beta} 
&=&
\frac{1}{n \cdot \partial} \varepsilon_{\mu \nu \alpha \beta}
n^{\alpha}
\cdot
e \sum_{j=1}^3 \vec{w}_j \; \bar{q}_j \gamma^{\beta} q_j.
\nonumber\\
&=&
e \sum_{j=1}^3 \vec{w}_j
\cdot \frac{1}{n \cdot \partial} \varepsilon_{\mu \nu \alpha \beta}
n^{\alpha} \bar{q}_j \gamma^{\beta} q_j
\nonumber\\
&\equiv&
e \sum_{j=1}^3 \vec{w}_j \Sigma_{j\;\mu\nu}.
\label{eqn:non-local-gene}
\eea
Here, one finds that $\Sigma_{j\;\mu\nu}$ describes 
the color-electric Dirac string singularity attached to 
the color-electric charge of $e\vec{w}_j$,  
which satisfies a similar quantization condition in integral form 
like (\ref{eqn:dirac-quantization}).
In this case, we have
\be
\int_S \Sigma_{j\; \mu \nu} d \sigma^{\mu\nu} = \pm 1,
\label{eqn:quantization-su3}
\ee
where $S$ is a certain surface which is pierced once by
the color-electric Dirac string.

\par
The color-magnetic charge of the monopole is 
defined by $g \vec{\epsilon}_i$, where $\vec{\epsilon}_i$ are the root 
vectors of the SU(3) algebra,
\be
\vec{\epsilon}_1=\left (-\frac{1}{2},\frac{\sqrt{3}}{2} \right ),\quad 
\vec{\epsilon}_2=\left(-\frac{1}{2},-\frac{\sqrt{3}}{2} \right ), \quad 
\vec{\epsilon}_3=\left (1,0 \right ),
\ee
where the labels 1, 2, 3 correspond to 
dual red (${}^{*\!}R$), dual blue (${}^{*\!}B$) and 
dual green (${}^{*\!}G$).  Here, `` ${}^{*}$ '' denotes dual.
Both the gauge coupling $e$ and the dual gauge coupling $g$ 
are related by the Dirac quantization condition $eg=4\pi$.
It might be worthwhile to remember that the relation of 
the root vector and the weight vector of the SU(3) algebra is given by
\bea
\vec{\epsilon}_i \cdot \vec{w}_j =
\frac{1}{2} \left (\begin{array}{ccc}
 0 & 1 &  -1 \\
 -1 &  0 & 1 \\
1 &  -1 &  0
\end{array}\right )
=
\frac{1}{2}\sum_{k=1}^3 \epsilon_{ijk}
\equiv
\frac{1}{2} m_{i j},
\label{eqn:charge-relation}
\eea
where $m_{ij}$ is an integer which takes 0 or $\pm 1$.

\par
The typical scale of the [U(1)]$^2$ DGL theory is determined by taking into 
account the dual Higgs mechanism as the U(1) DGL theory.
By inserting $\chi_i = \left ( v + \phi_i/\sqrt{2} \right ) e^{i\eta_i}$ 
(where $\phi_i, \eta_i \in \Re$) into the [U(1)]$^2$ DGL Lagrangian 
(\ref{eqn:DGL}), we get
\bea
{\cal L}_{\rm [U(1)]^2\;DGL} 
&=&
-\frac{1}{4} {}^{*\!}\vec{F}_{\mu \nu}^2 (\vec{B}',\vec{j})
+\frac{1}{2} m_B^2 \vec{B}_{\mu}^{'2} 
+
\sum_{i=1}^3 \frac{1}{2} 
\left [ (\partial_{\mu}\phi_i)^2 -m_\chi^2 \phi_i^2 \right ]
\nonumber\\
&&
+\sum_{i=1}^3 
\left [ 
g^2 (\vec{\epsilon}_i\cdot \vec{B}_{\mu}')^2 
\left ( \sqrt{2} v \phi_i + \frac{\phi_i^2}{2} \right )
-\lambda \left (\sqrt{2}v \phi_i^3 + \frac{\phi_i^4}{4} \right )
\right ],
\eea
where the phase of the monopole field $\eta_i$ 
is absorbed into the dual gauge field $\vec{B}_{\mu}'$, 
as $\vec{\epsilon}_i \cdot \vec{B}_{\mu}' =
\vec{\epsilon}_i \cdot \vec{B}_{\mu} +\partial_{\mu}\eta_i /g$,
and accordingly
the dual gauge field and the monopole field acquire the masses, 
$m_B= \sqrt{3}gv$, $m_\chi = 2 \sqrt{\lambda}v $, respectively.
The GL parameter is then given by
\bea
\kappa \equiv \frac{m_B^{-1}}{m_\chi^{-1}} 
= \frac{2\sqrt{\lambda}}{\sqrt{3}g}.
\eea
As explained in U(1) DGL theory, $\kappa=1$ is the case of special interest, 
the Bogomol'nyi limit \cite{Chernodub:1999xi,Koma:2000wn}.

\subsection{Various representations of the dual gauge field}

\par
The color-electric charge of the quark is given by three components as 
$R$, $B$ and $G$, which is spanned by the weight vector of SU(3) 
algebra.
The color-magnetic charge of the monopole is defined by components as
${}^{*\!}R$, ${}^{*\!}B$ and ${}^{*\!}G$,  which is spanned by  
the root vector of SU(3) algebra. 
Now, we are interested in the color-singlet state corresponding 
to the meson and the baryon, which should be invariant under the exchange 
of the color charges. Hence, it is important to pay attention to the 
Weyl symmetry in the DGL theory.
However, since the dual gauge field $\vec{B}_{\mu}$ which connects 
the color-electric charge and the color-magnetic 
charge has only two components in the sense of Cartan decomposition, 
and accordingly, the independent color-electric flux has two 
components, we cannot observe the Weyl symmetric structure 
in the color-electric flux tube itself. 
This fact makes it difficult to see the Weyl invariant structure of 
hadronic states.
In order to investigate the Weyl symmetric structure of the
flux tube in the DGL theory, 
it would be favorable to represent the dual gauge field in a 
Weyl symmetric way.

\par
In this subsection, we write the [U(1)]$^2$ DGL Lagrangian in various 
representations of the dual gauge field, among which the Weyl symmetric 
representation of the dual gauge field is also discussed.
We first pay attention to the original Cartan representation of the 
dual gauge field with two components.
Next, we will discuss other two possible representations of the dual gauge 
field, the color-electric representation and the color-magnetic 
representation, which are achieved by spanning the dual gauge field 
with the weight vector and the root vector, respectively.

\subsubsection{Cartan 3-8 representation}

The two-component dual gauge field $\vec{B}_{\mu}$ can be written as
\be
\vec{B}_{\mu} \equiv 
\frac{1}{g}
\left ( B^3_{\mu}, 
\frac{B^8_{\mu}}{\sqrt{3}}\right ) .
\ee
The factor $\sqrt{3}$ is to make 3- and 8- components symmetric.
The dual field strength tensor becomes
\bea
{}^{*\!} \vec{F}_{\mu\nu}
&=&
\frac{1}{g}
\Biggl (
\partial_{\mu}B_{\nu}^3 - \partial_{\nu}B_{\mu}^3
-2\pi (\Sigma_{1\;\mu\nu}-\Sigma_{2\;\mu\nu})
, \nonumber\\
&&
\frac{1}{\sqrt{3}} 
\left \{
\partial_{\mu}B_{\nu}^8 - \partial_{\nu}B_{\mu}^8
-2\pi (\Sigma_{1\;\mu\nu}+\Sigma_{2\;\mu\nu} -2 \Sigma_{3\;\mu\nu})
\right \}
\Biggr ) \nonumber\\
&\equiv&
\frac{1}{g}
\left ( {}^{*\!} F_{\mu\nu}^3, 
\frac{{}^{*\!} F_{\mu\nu}^8}{\sqrt{3}}  \right ),
\eea
where we have used $eg=4\pi$ to get the factor $2\pi$ in front of 
$\Sigma_{j\;\mu\nu}$.
The [U(1)]$^2$ DGL Lagrangian is written as
\bea
{\cal L}_{\rm [U(1)]^2 DGL}^{3-8} &=&
-\frac{1}{4 g^2} ({}^{*\!} F_{\mu\nu}^3)^2
-\frac{1}{12 g^2} ({}^{*\!} F_{\mu\nu}^8)^2
\nonumber\\
&&
+
\left | \left (\partial_{\mu}+
i\frac{1}{2}\left ( -B_{\mu}^3 + B_{\mu}^8 \right )
\right )\chi_{1} \right |^2
-\lambda \left ( \left |\chi_{1} \right |^2-v^2 \right )^2 \nonumber\\
&&
+
\left | \left (\partial_{\mu}+
i\frac{1}{2}\left ( -B_{\mu}^3 - B_{\mu}^8 \right )
\right )\chi_{2} \right |^2
-\lambda \left ( \left |\chi_{2} \right |^2-v^2 \right )^2 \nonumber\\
&&
+
\left | \left (\partial_{\mu}+
i B_{\mu}^3 
\right )\chi_{3} \right |^2
-\lambda \left ( \left |\chi_{3} \right |^2-v^2 \right )^2.
\label{eqn:DGL-3-8}
\eea
Note that the Lagrangian (\ref{eqn:DGL-3-8}) is invariant 
under the [U(1)]$^2$ dual gauge transformation,
\bea
&&
\chi_{i} \to  \chi_{i} e^{if_{i}}, \quad 
\chi^*_{i} \to \chi^*_{i} e^{-if_{i}} \quad\quad (i=1,2,3),\nonumber\\
&&
(B_{\mu}^3,\; B_{\mu}^8)
\to
\left ( B_{\mu}^3 - \partial_{\mu}f_3,\;
B_{\mu}^8 - (\partial_{\mu}f_1 -\partial_{\mu}f_2)
\right ),
\label{eqn:gauge-sym-3-8}
\eea
where the phases $f_i$ fulfill the constraint
$\sum_{i=1}^3 f_{i}= 0$ \cite{Suzuki:1988yq,Suganuma:1995ps}.

\par
The field equations are given by
\bea
\frac{1}{g^2}
\partial^{\nu} {}^{*\!}F_{\mu \nu}^3
&=&
+
\frac{i}{2} \left (
\chi_1^* \partial_{\mu}\chi_1 - \chi_1 \partial_{\mu}\chi_1^*
\right )
-\frac{1}{2}\left (-B_{\mu}^3 + B_{\mu}^8 \right )\chi_1^* \chi_1
\nonumber\\
&&
+
\frac{i}{2} \left (
\chi_2^* \partial_{\mu}\chi_2 - \chi_2 \partial_{\mu}\chi_2^*
\right )
-\frac{1}{2}\left (-B_{\mu}^3 - B_{\mu}^8 \right )\chi_2^* \chi_2
\nonumber\\
&&
- i 
\left ( \chi_3^* \partial_{\mu}\chi_3 - \chi_3 \partial_{\mu}\chi_3^*
\right )
+2 B_{\mu}^3 \chi_3^* \chi_3,
\eea
\bea
\frac{1}{3g^2}
\partial^{\nu} {}^{*\!}F_{\mu \nu}^8
&=&
-\frac{i}{2} \left (
\chi_1^* \partial_{\mu}\chi_1 - \chi_1 \partial_{\mu}\chi_1^*
\right )
+\frac{1}{2}\left (-B_{\mu}^3 + B_{\mu}^8 \right )\chi_1^* \chi_1
\nonumber\\
&&
+
\frac{i}{2} \left (
\chi_2^* \partial_{\mu}\chi_2 - \chi_2 \partial_{\mu}\chi_2^*
\right )
-\frac{1}{2}\left (-B_{\mu}^3 - B_{\mu}^8 \right )\chi_2^* \chi_2,
\eea
\bea
\left ( \partial_{\mu} +
\frac{i}{2} \left (-B_{\mu}^3 + B_{\mu}^8 \right ) 
\right )^2 \chi_1 
&=&
-2 \lambda \chi_1 \left ( \chi_1^* \chi_1-v^2 \right ),
\\
\left ( \partial_{\mu} +
\frac{i}{2} \left (-B_{\mu}^3 - B_{\mu}^8 \right ) 
\right )^2 \chi_2
&=&
-2 \lambda \chi_2 \left ( \chi_2^* \chi_2-v^2 \right ),
\\
\left ( \partial_{\mu} + i B_{\mu}^3 \right )^2 \chi_3
&=&
-2 \lambda \chi_3 \left ( \chi_3^* \chi_3 - v^2 \right ).
\eea
From these field equations, we find the boundary conditions : 
If ${}^{*\!}F_{\mu \nu}^3$ and ${}^{*\!}F_{\mu \nu}^8$ have 
a non-vanishing nonlocal term $\Sigma_{j\;\mu\nu}$, the 
dual gauge field $B_{\mu}^3$ and $B_{\mu}^8$ also have the singular part.
At the place where the dual gauge field is singular, the monopole 
field is required to disappear.
At large distance from the singularity,
the monopole field $\chi_i$ approaches the vacuum expectation value $v$
and the dual gauge field asymptotically vanishes, $B_{\mu}^3=B_{\mu}^8=0$.
These field equations are to be solved by using the dual lattice formulation,  
and one will find that these boundary conditions are realized.

\subsubsection{Color-electric representation}

The dual gauge field can be expressed by using the weight vector $\vec{w}_j$, 
where the label $j=1,2,3$ corresponds to the color-electric charge,
$R$, $B$ and $G$.
In this sense, we call this the color-electric 
representation of the dual gauge field, which is defined by 
\bea
\vec{B}_{\mu} 
\equiv
\sqrt{\frac{2}{g_e^2}}
\sum_{j=1}^3 
\vec{w}_{j} B_{j\;\mu}^e,
\label{eqn:b-weyl-elect}
\eea
where 
\be
g_e \equiv \frac{3}{\sqrt{2}}g,\quad
B_{j\;\mu}^e \equiv \sqrt{2} g^e \vec{w}_{j} \cdot \vec{B}_{\mu}.
\ee
Note that now the dual gauge field is written as a three-component
field, however all of them are not independent since
$\sum_{j=1}^3 B_{j\;\mu}^e = 0$.
The dual field strength tensor has the form
\bea
{}^{*\!}\vec{F}_{\mu\nu}
=
\sqrt{\frac{2}{g_e^2}}
\sum_{j=1}^3 \vec{w}_j
\left (
\partial_{\mu}B_{j \; \nu}^e -\partial_{\nu} B_{j\;\mu}^e
- 2 \pi \Sigma_{j\; \mu \nu}
\right ),
\eea
where $eg=4\pi$ is used.
Then, we get the Lagrangian
\bea
{\cal L}_{\rm [U(1)]^2 DGL}^{\rm electric} 
&=&
-\frac{1}{4 g_e^2 }
\sum_{j=1}^3
\left ( {}^{*\!} F_{j\;\mu\nu}^e \right )^2 \nonumber\\
&&
+\sum_{i=1}^3
\left [ \left | \left (\partial_{\mu} 
+ i \frac{1}{3}\sum_{j=1}^3 m_{ij} B_{j\;\mu}^e 
\right )\chi_{i} \right |^2
-\lambda \left ( \left |\chi_{i} \right |^2-v^2 \right )^2 \right ],
\label{eqn:DGL-elect}
\eea
where
\be
{}^{*\!}F_{j\;\mu\nu}^e 
\equiv
 \partial_{\mu}B_{j\;\nu}^e - \partial_{\nu}B_{j\;\mu}^e
- 2 \pi \left ( 2 \Sigma_{j\; \mu \nu} - \sum_{k=1}^3 m_{jk}^2
\Sigma_{k\; \mu \nu} \right ).
\ee
Here, we have used the relations
\be
g\vec{\epsilon}_{i}{\cdot}\vec{B}_{\mu} 
=
g\vec{\epsilon}_{i}{\cdot} \sqrt{\frac{2}{g_e^2}}
\sum_{j=1}^3 \vec{w}_{j} B_{j\;\mu}^e
=
\frac{1}{3} \sum_{j=1}^3 m_{ij} B_{j\;\mu}^e,
\ee
\be
\vec{\epsilon}_i = \vec{w}_j - \vec{w}_k
\quad (i,j,k : cyclic).
\ee
Apparently, the Lagrangian (\ref{eqn:DGL-elect}) is invariant 
under the [U(1)]$^3$ dual gauge transformation, which is defined by
\bea
&&\chi_{i} \to  \chi_{i} e^{if_{i}},\quad
\chi^*_{i} \to \chi^*_{i} e^{-if_{i}},\nonumber\\
&&
B_{j\;\mu}^e \to B_{j\;\mu}^e + \sum_{i=1}^3 m_{j i} \partial_{\mu} f_i,
\label{eqn:gauge-sym-weyl-e}
\eea
where $i,j=1,2,3$.
However, this does not mean an increase of the gauge degrees of freedom
because of the constraint $\sum_{j=1}^3 B_{j\;\mu}^e =0$.

\par
The field equations for $j = 1,2,3$ and $i=1,2,3$ are given by
\bea
&&
\frac{1}{g_e^2} \partial^{\nu} {}^{* \!} F_{j\; \mu \nu}^e 
=
\sum_{i=1}^3 m_{ij} 
\left [ 
-\frac{i}{3} 
\left ( 
\chi_{i}^{*}\partial_{\mu} \chi_{i} - 
\chi_{i}\partial_{\mu} \chi_{i}^{*}
\right )
+
2 \sum_{k=1}^3
m_{ik} B_{k\;\mu}^e \chi_{i}^{*}\chi_{i} 
\right ],
\label{eqn:feq-1-e}\\
&&
\left ( \partial_{\mu} 
+i\frac{1}{3} \sum_{j=1}^3 m_{ij} B_{j\;\mu}^e 
\right )^2 \chi_i
=
- 2 \lambda \chi_{i} ( \chi_{i}^{*} \chi_{i} - v^2).
\label{eqn:feq-2-e}
\eea
We find that each field equation has U(1) structure, apart from the 
matrix structure in labels $i$ and $j$.
The boundary condition is given by a similar discussion as in the
Cartan representation of the dual gauge field.
The main difference is that the dual gauge field represented here
experiences the color-electric Dirac string singularity in a  Weyl
symmetric way.
The dual lattice formulation will make this situation clear.

\subsubsection{Color-magnetic representation}

The dual gauge field can also be spanned by using the 
root vector $\vec{\epsilon}_i$,
where the label $i=1,2,3$ corresponds to the monopole charge, 
${}^{*\!}R$, ${}^{*\!}B$ and ${}^{*\!}G$.
In this sense, we call this the color-magnetic representation of 
the dual gauge field \cite{Koma:2000wn}, defined by
\bea
\vec{B}_{\mu} 
\equiv
\sqrt{\frac{2}{3 g_m^2}} 
\sum_{i=1}^3 \vec{\epsilon}_{i} B_{i\;\mu}^m,
\label{eqn:b-weyl-mag}
\eea
where
\be
g_m \equiv \sqrt{\frac{3}{2}}g, \quad \quad
B_{i\;\mu}^m \equiv \sqrt{\frac{2}{3}} g_m 
\vec{\epsilon}_{i} \cdot \vec{B}_{\mu} .
\ee
Note that all $B_{i\;\mu}^m$ are not independent 
since $\sum_{i=1}^3 B_{i\;\mu}^m=0$.
The dual field strength tensor is written as
\bea
{}^{*\!}\vec{F}_{\mu\nu}
=
\sqrt{\frac{2}{3 g_m^2}}
\sum_{i=1}^3 \vec{\epsilon}_i
\left (
\partial_{\mu}B_{i \; \nu}^m -\partial_{\nu} B_{i\;\mu}^m
- 2 \pi \sum_{j=1}^3 m_{ij} \Sigma_{j\; \mu \nu}
\right ),
\eea
where we use $eg=4\pi$. 
Hence, the Lagrangian with the color-magnetic representation of the 
dual gauge field is given by
\bea
{\cal L}_{\rm [U(1)]^2 DGL}^{\rm magnetic}
&=&
\sum_{i=1}^3
\Biggl [
- \frac{1}{4 g_m^2} 
\left ( {}^{*\!}F_{i\;\mu\nu}^m \right )^2 
+ \left | \left (\partial_{\mu} + i B_{i\;\mu}^m \right ) \chi_{i} \right |^2
- \lambda \left ( \left |\chi_{i} \right |^2-v^2 \right )^2
\Biggr ],
\label{eqn:dgl-weyl-mag}
\eea
where
\be
{}^{*\!}F_{i\;\mu\nu}^m
\equiv
 \partial_{\mu} B_{i\;\nu}^m -\partial_{\nu} B_{i\;\mu}^m
- 2 \pi \sum_{j=1}^3 m_{ij} \Sigma_{j\; \mu \nu}.
\ee
Here, we have used the relations
\be
\vec{w}_i = -\frac{1}{3}
\left (\vec{\epsilon}_j - \vec{\epsilon}_k \right ) 
\quad (i,j,k : cyclic) .
\ee

\par
Since the Lagrangian (\ref{eqn:dgl-weyl-mag}) has a quite similar 
form as the U(1) DGL theory, except for the labels $i$ and $j$,
one finds that the dual gauge symmetry becomes very easy 
to observe, 
\bea
&&\chi_{i} \to  \chi_{i} e^{if_{i}},\quad
\chi^*_{i} \to \chi^*_{i} e^{-if_{i}},\quad
B_{i\;\mu}^m \to B_{i\;\mu}^m - \partial_{\mu} f_i  \quad\quad ( i=1,2,3),
\label{eqn:gauge-sym-weyl-m}
\eea
and accordingly the Lagrangian (\ref{eqn:dgl-weyl-mag}) has the 
extended dual gauge symmetry [U(1)]$^3$ with a 
constraint $\sum_{i=1}^3 B_{i\;\mu}^m = 0$.
This is the same as in  the color-electric representation of the 
dual gauge field.

\par
The field equations for $i=1,2,3$ have the form
\bea
&&
\frac{1}{g_m^2} \partial^{\nu} {}^{* \!} F_{i\; \mu \nu}^m 
=
-i \left ( 
\chi_{i}^{*}\partial_{\mu} \chi_{i} - 
\chi_{i}\partial_{\mu} \chi_{i}^{*}
\right )
+
2 B_{i\;\mu}^m \chi_{i}^{*}\chi_{i},
\label{eqn:feq-1-m}\\
&&
\left ( \partial_{\mu} +i B_{i\;\mu}^m \right )^2 \chi_i
=
- 2 \lambda \chi_{i} ( \chi_{i}^{*} \chi_{i} - v^2),
\label{eqn:feq-2-m}
\eea
which is exactly the same as the field equation in the U(1) DGL theory, 
replicated with respect to the index $i$.
In this sense, the boundary conditions can be taken similarly as 
the U(1) case.
Therefore, the color-magnetic representation of the dual gauge field
is particularly simple as compared with other representations.

\subsection{The solution}

\par
In order to solve the field equation with various representations of 
the dual gauge field, we adopt the dual lattice formulation  
with the U(1) DGL theory, but extended to more degrees of freedom. 
In this subsection, we first investigate the mesonic flux tube, and next 
the baryonic flux tube. We use the words `` mesonic '' or `` baryonic '' 
to distinguish the real color-singlet hadron from the classical state 
that we deal with in this paper. 
For instance, if we want to obtain real meson or baryon state, we 
need to consider the quantum state given by 
\bea
|meson\rangle &=& \frac{1}{\sqrt{3}} 
\left ( |R\bar{R}\rangle + |B\bar{B}\rangle
+|G\bar{G}\rangle \right ),
\nonumber\\
|baryon\rangle &=& \frac{1}{\sqrt{6}}
\left ( |RBG\rangle + |BGR\rangle +|GRB\rangle 
-|RGB\rangle - |GBR\rangle - |BRG\rangle  \right ),\nonumber
\eea
where $R\bar{R}$ denotes $R-\bar{R}$ flux tube, and so forth.
In the classical solution, we can only treat a piece of these states.
However, even then it is necessary  to pay attention to the Weyl symmetry, 
since all states can be reduced to the same classical state for the meson 
and the baryon, respectively.

\subsubsection{Mesonic flux tube ($q$-$\bar{q}$ system)}

Since the three types of the color-electric charge
are represented by non-vanishing plaquettes $\hat{\Sigma}_{x,1\;\mu\nu}$,
$\hat{\Sigma}_{x,2\;\mu\nu}$ and $\hat{\Sigma}_{x,3\;\mu\nu}$, 
the mesonic state corresponding to $|R\bar{R}\rangle$, 
$|B\bar{B}\rangle$ and $|G\bar{G}\rangle$ are given 
by some stacks of connected plaquettes of each color.
For example, if we want to consider the straight $R-\bar{R}$ 
flux-tube system,
all we have to do is to put only  one of the color-electric Dirac
string plaquette $\hat{\Sigma}_{x,1\;\mu\nu}\ne 0$ like 
Fig.~\ref{fig:sigular-pla}(a), 
whereas $\hat{\Sigma}_{x,2\;\mu\nu}=\hat{\Sigma}_{x,3\;\mu\nu}=0$ for all over the
three dimensional space.
For the $B-\bar{B}$ flux-tube system, we set $\hat{\Sigma}_{x,2\;\mu\nu}\ne 0$
and $\hat{\Sigma}_{x,3\;\mu\nu}=\hat{\Sigma}_{x,1\;\mu\nu}=0$,
for the $G-\bar{G}$ flux-tube system, $\hat{\Sigma}_{x,3\;\mu\nu}\ne 0$ and
$\hat{\Sigma}_{x,1\;\mu\nu}=\hat{\Sigma}_{x,2\;\mu\nu}=0$.

\par
\begin{figure}[t]
\centerline{\epsfxsize = 9cm\epsfbox{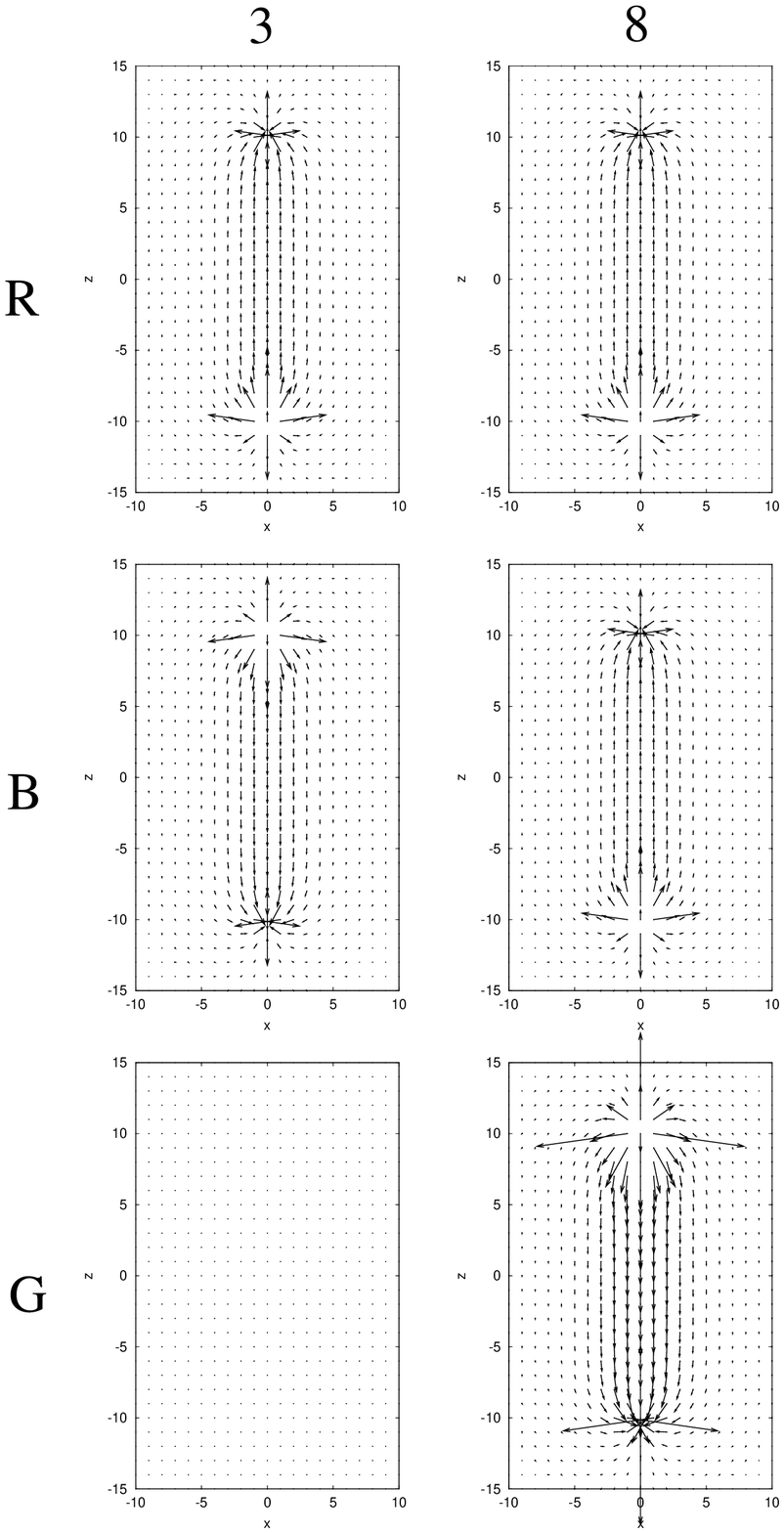}}
\caption{The profiles of the color-electric field in the Cartan 
representation for 3- (left) and 8- (right)
components in the $R-\bar{R}$ (upper), the $B-\bar{B}$ (middle), and
 the $G-\bar{G}$ (lower) systems in the $x$-$z$ plane at $y=0$.
The quark and the antiquark are placed at 
$(x,y,z)=(0,0,-10)$ and $(0,0,10)$, respectively.}
\label{fig:38-meson-ele}
\end{figure}
\begin{figure}[t]
\centerline{\epsfxsize = 12cm\epsfbox{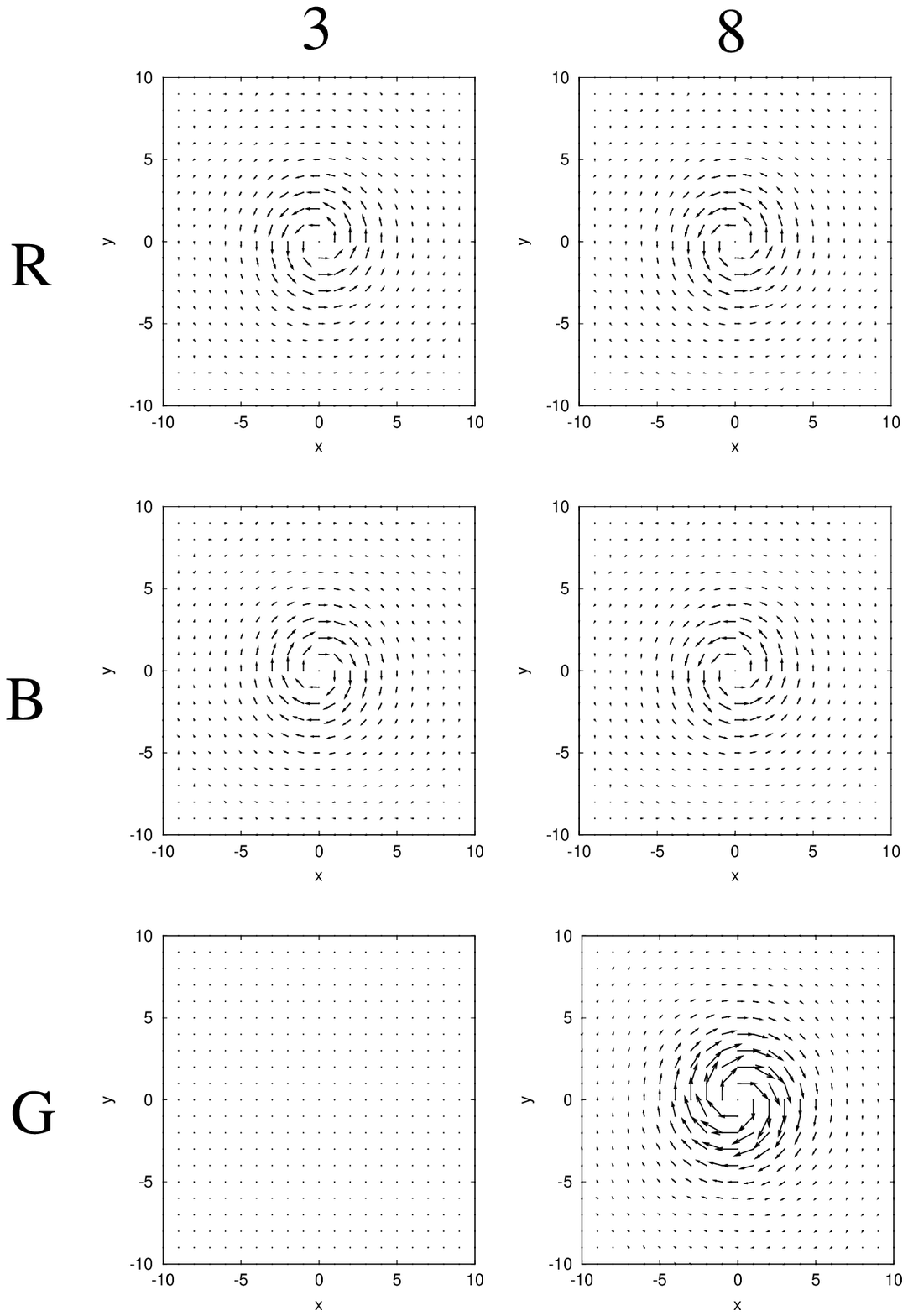}}
\caption{The profiles of the color-magnetic current in the Cartan 
representation for 3- (left) and 8- (right)
components in  the $R-\bar{R}$ (upper),  the $B-\bar{B}$ (middle), and 
the $G-\bar{G}$ (lower) systems in the $x$-$y$ plane at $z=0$.
The quark and the antiquark are placed at 
$(x,y,z)=(0,0,-10)$ and $(0,0,10)$, respectively.}
\label{fig:38-meson-mag}
\end{figure}
\begin{figure}[t]
\centerline{\epsfxsize = 13cm\epsfbox{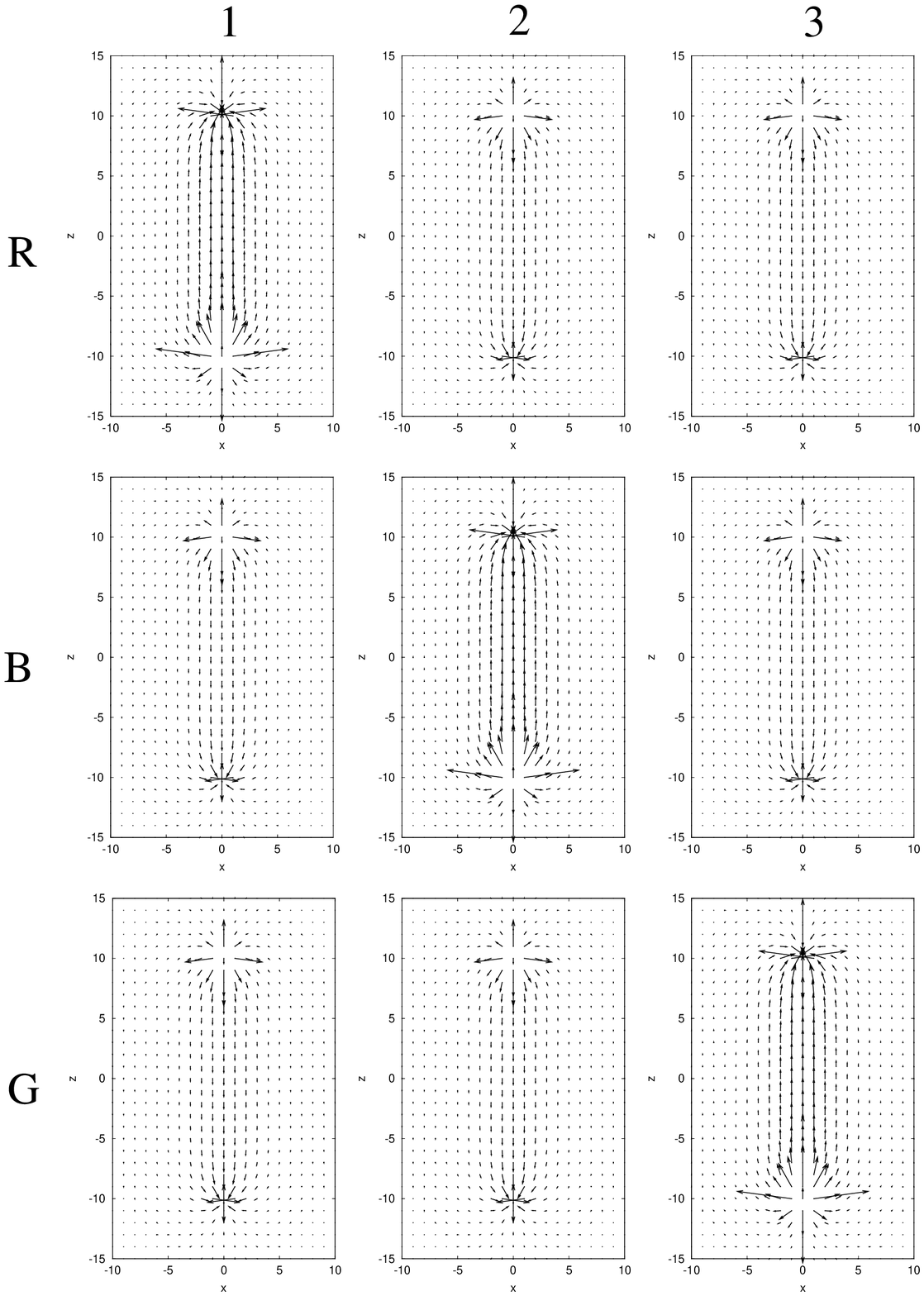}}
\caption{The profiles of the color-electric field in the color-electric
representation, expressed on the weight
vectors of the SU(3) algebra, $\vec{w}_1$ (left),
$\vec{w}_2$ (center), and $\vec{w}_3$ (right)
in  the $R-\bar{R}$ (upper), the $B-\bar{B}$ (middle), and
 the $G-\bar{G}$ (lower) systems in the $x$-$z$ plane at $y=0$.
The quark and the antiquark are placed at 
$(x,y,z)=(0,0,-10)$ and $(0,0,10)$, respectively.}
\label{fig:ele-meson-ele}
\end{figure}
\begin{figure}[t]
\centerline{\epsfxsize = 15cm\epsfbox{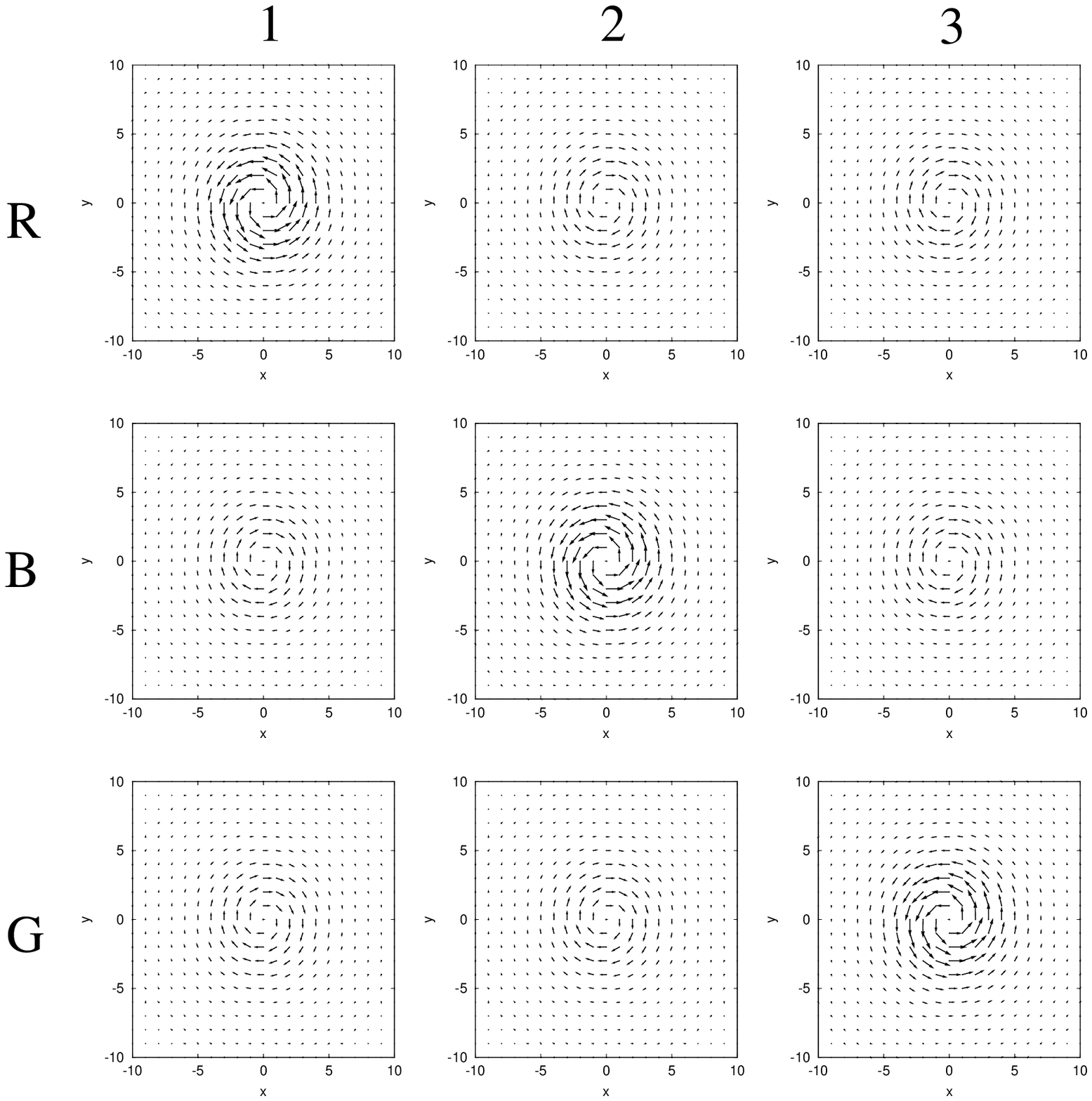}}
\caption{The profiles of the color-magnetic current in the color-electric
representation, expressed on the weight
vectors of the SU(3) algebra, $\vec{w}_1$ (left),
$\vec{w}_2$ (center), and $\vec{w}_3$ (right)
in the $R-\bar{R}$ (upper), the $B-\bar{B}$ (middle), and
 the $G-\bar{G}$ (lower) systems in the $x$-$y$ plane at $z=0$.
The quark and the antiquark are placed at 
$(x,y,z)=(0,0,-10)$ and $(0,0,10)$, respectively.}
\label{fig:ele-meson-mag}
\end{figure}
\begin{figure}[t]
\centerline{\epsfxsize = 13cm\epsfbox{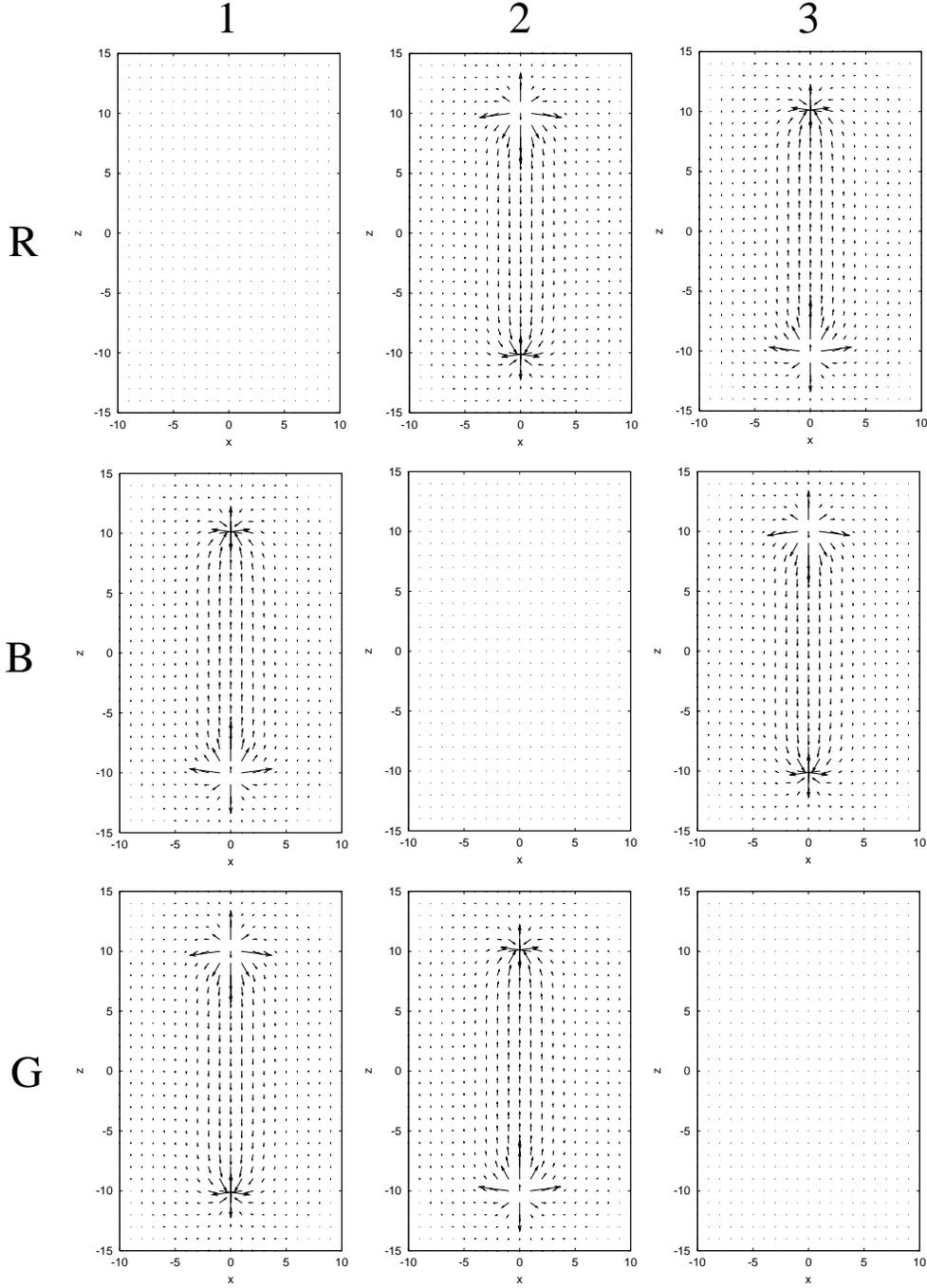}}
\caption{The profiles of the color-electric field in the
color-magnetic representation, expressed on the root
vectors of the SU(3) algebra, $\vec{\epsilon}_1$ (left),
$\vec{\epsilon}_2$ (center), and $\vec{\epsilon}_3$ (right)
in  the $R-\bar{R}$ (upper), the $B-\bar{B}$ (middle), and
 the $G-\bar{G}$ (lower) systems in the $x$-$z$ plane at $y=0$.
The quark and the antiquark are placed at 
$(x,y,z)=(0,0,-10)$ and $(0,0,10)$, respectively.}
\label{fig:mag-meson-ele}
\end{figure}
\begin{figure}[t]
\centerline{\epsfxsize = 15cm\epsfbox{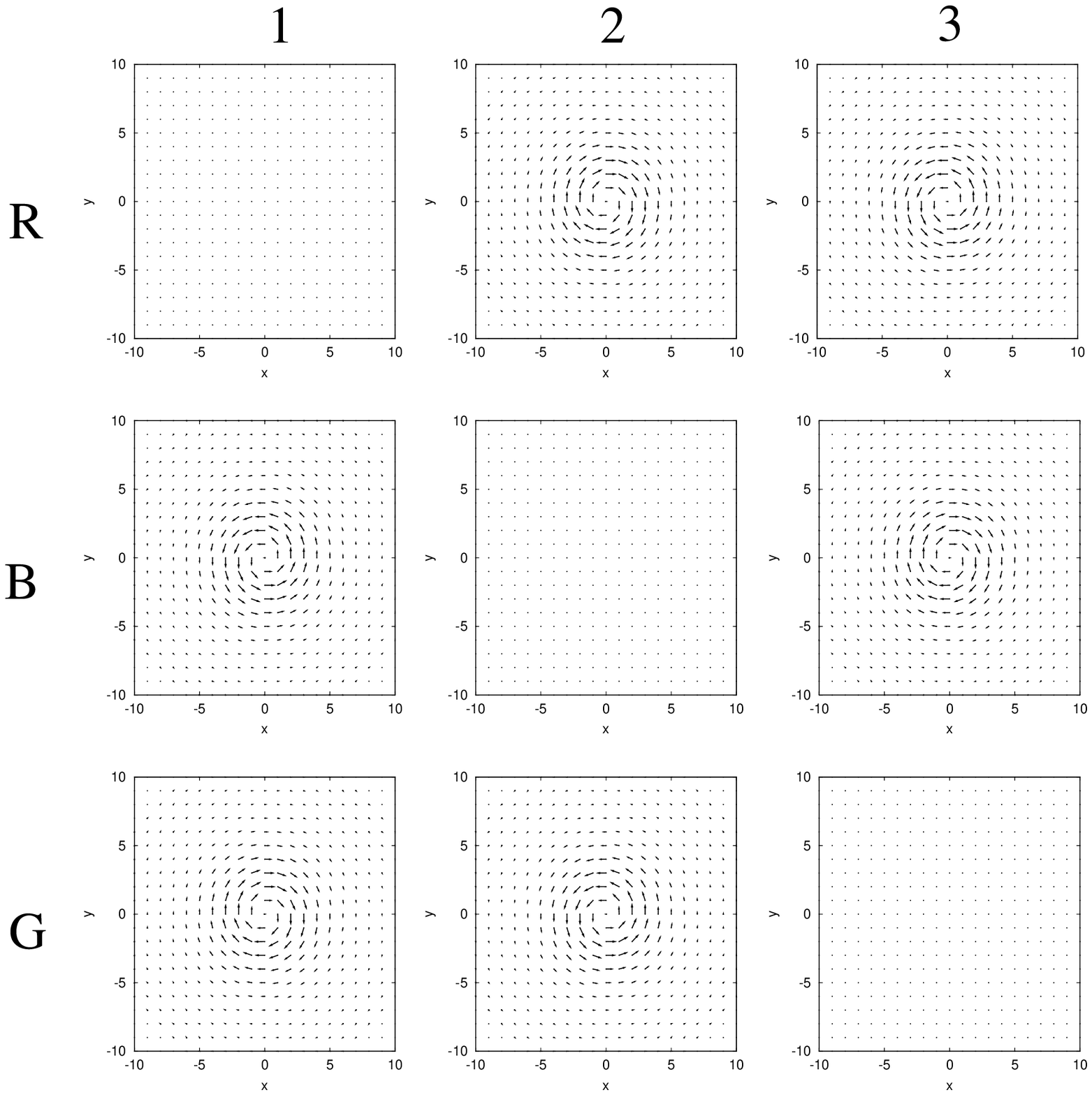}}
\caption{The profiles of the color-magnetic current in the color-magnetic
representation, expressed on the root
vectors of the SU(3) algebra, $\vec{\epsilon}_1$ (left),
$\vec{\epsilon}_2$ (center), and $\vec{\epsilon}_3$ (right)
in the $R-\bar{R}$ (upper), the $B-\bar{B}$ (middle), and
 the $G-\bar{G}$ (lower) systems in the $x$-$y$ plane at $z=0$.
The quark and the antiquark are placed at 
$(x,y,z)=(0,0,-10)$ and $(0,0,10)$, respectively.}
\label{fig:mag-meson-mag}
\end{figure}

In Figs.~\ref{fig:38-meson-ele}-\ref{fig:mag-meson-mag}, we show 
the profiles of the color-electric field 
and corresponding monopole current of 
$R-\bar{R}$, $B-\bar{B}$, and $G-\bar{G}$ flux-tube system for 
various representation of the dual gauge field, the Cartan representation,
the color-electric representation, and the color-magnetic representation, 
respectively.
We find that the last two representations enable us to see
the Weyl symmetric structure of the flux tube.
The Dirac string structures {\em in the dual gauge field} with 
various representations is summarized schematically in Table.~1.

\par
\begin{figure}[t]
\centerline{\epsfxsize = 15cm\epsfbox{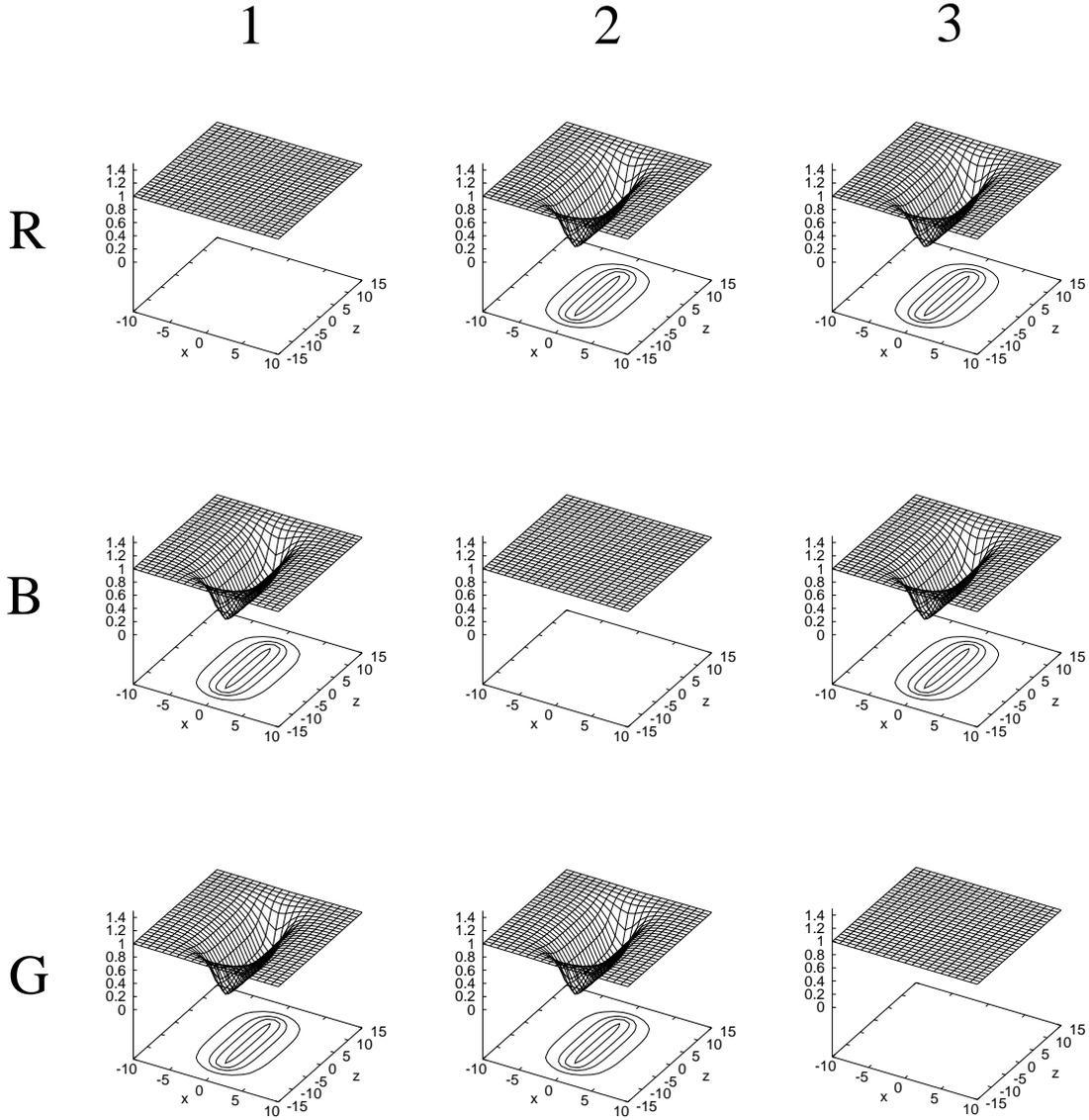}}
\caption{The profiles of the monopole field of $|\chi_1|$ (left),
$|\chi_2|$ (center), and $|\chi_3|$ (right)
in the $R-\bar{R}$ (upper), the $B-\bar{B}$ (middle), and
the $G-\bar{G}$ (lower) systems in the $x$-$z$ plane at $y=0$.
The quark and the antiquark are placed at 
$(x,y,z)=(0,0,-10)$ and $(0,0,10)$, respectively.}
\label{fig:meson-higgs}
\end{figure}
\begin{figure}[t]
\centerline{\epsfxsize = 10cm\epsfbox{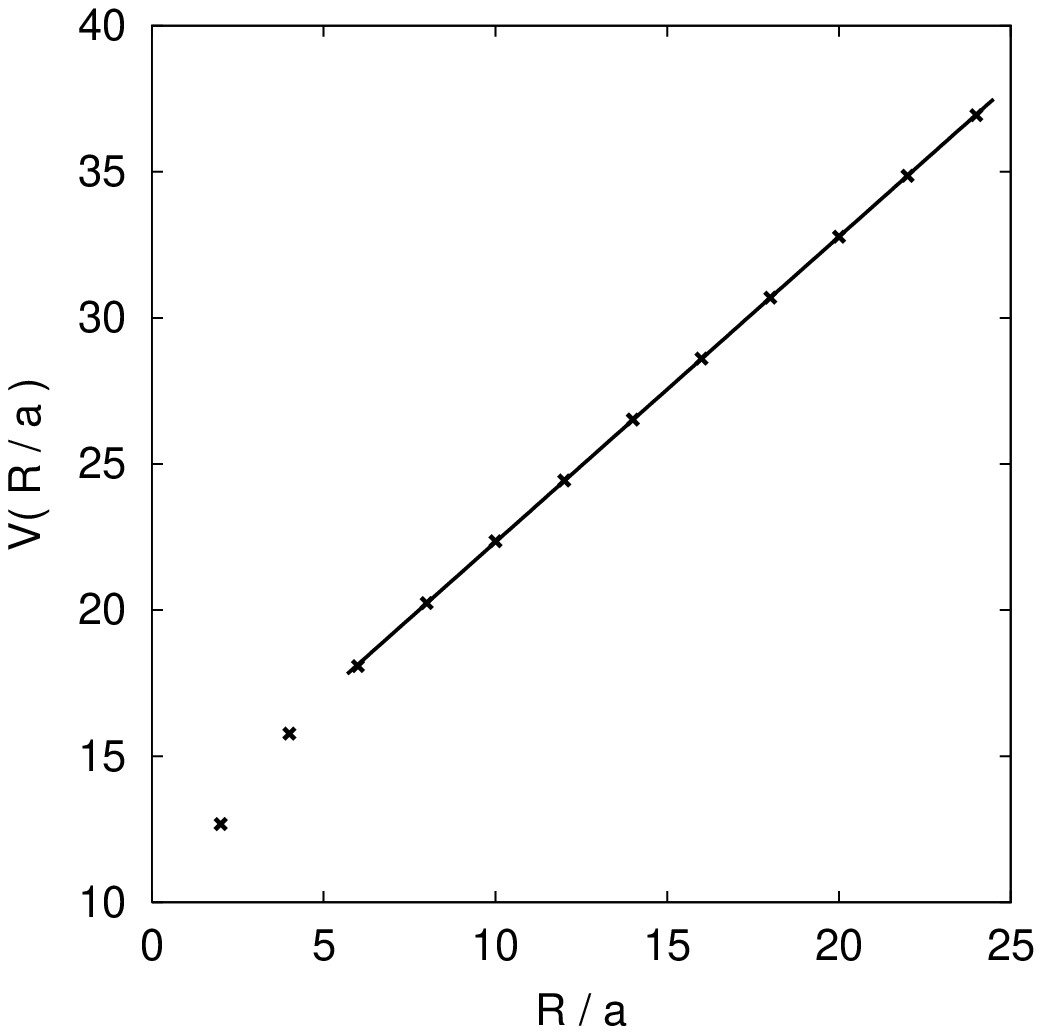}}
\caption{The quark-antiquark potential in the [U(1)]$^2$ DGL theory, 
where $R/a$ denotes the $q$-$\bar{q}$ distance.
The parameter set is taken as 
$\beta = 1$, $\hat{m}_B = \hat{m}_\chi =0.5$.}
\label{fig:meson-pot}
\end{figure}

The profile of the monopole field is shown in Fig.~\ref{fig:meson-higgs}. 
One finds that this does not depend on 
the choice of the representation of the dual gauge field, since the 
monopole field is defined on the SU(3) root vector. 
That is the reason why this distribution is similar to the 
color-electric field in the color-magnetic representation of 
the dual gauge field.
The inter-quark potential is shown in Fig.~\ref{fig:meson-pot}, 
which, of course, does not depend on the representation. 
The parameter set used here is the same as in the  U(1) case.
We took $\beta \equiv 1/g^2 =1$, $\hat{m}_B=\hat{m}_\chi=0.5$. 
This set is simply to see the behavior of the profiles and to compare
the string tension of the potential with the analytical value 
in the Bogomol'nyi limit, 
$\sigma^L = 4\pi v^2 \cdot a^2 = 4 \beta \pi \hat{m}_B^2/3$ 
\cite{Chernodub:1999xi,Koma:2000wn}.
One finds that the analytical string tension is reproduced by the numerical 
potential in Fig.~\ref{fig:meson-pot}.
In order to get quantitatively realistic
results, we need more information about the 
parameter set of U(1)$^2$ DGL theory from QCD.

\par
It is worth noting that in the mesonic case,
we can reduce the [U(1)]$^2$ DGL theory to the U(1) DGL 
theory \cite{Ichie:1996jr}.
Let us see this in the $R-\bar{R}$ system with the Cartan representation 
of the dual gauge field, as an example. 
Other systems and other representations can be treated similarly.
Here, we already know the profiles of the color-electric flux tube and 
the contribution of the dual gauge field and the monopole field 
as shown in Figs.~\ref{fig:38-meson-ele} and~\ref{fig:meson-higgs}.
Thus, one can take $B_{\mu}^3 = B_{\mu}^8 \equiv B_{\mu}$ and
$\chi_1 = v $, $\chi_2 \equiv \chi^*$, $\chi_3 \equiv \chi$.
The [U(1)]$^2$ DGL Lagrangian (\ref{eqn:DGL-3-8}) is reduced to 
the form,
\bea
{\cal L}_{\rm [U(1)]^2 DGL}^{3-8} &=&
-\frac{1}{3 g^2} \left ( \partial_{\mu} B_{\nu}-
\partial_{\nu} B_{\mu} -2 \pi \Sigma_{1\;\mu\nu} \right )^2
\nonumber\\
&&
+ 2 \left [
\left | \left (\partial_{\mu}+ i B_{\mu}\right )\chi \right |^2
-\lambda \left ( \left |\chi \right |^2-v^2 \right )^2
\right ].
\eea
The redefinitions of the couplings and the fields
\be
g \equiv \frac{2}{\sqrt{3}}\hat{g},\quad
\lambda \equiv 2\hat{\lambda},\quad
v \equiv \frac{1}{\sqrt{2}} \hat{v},\quad
B_{\mu} \to \hat{g} B_{\mu},\quad
\chi \to \frac{\chi}{\sqrt{2}},\quad
\ee
lead to the Lagrangian of U(1) DGL theory as is given in 
(\ref{eqn:DAHM-lagrangian}).

\subsubsection{Baryonic flux tube ($q$-$q$-$q$ system)}

We solve the field equations in the presence of three types of the 
color-electric charges.
Since these color-electric charges are defined in the weight 
vector diagram of SU(3) algebra, and the color-electric Dirac strings
which are attached to these charges carry the same quantity, 
respectively, these Dirac strings can join at a 
certain point to cancel each other ($\sum_{j=1}^3 e \vec{w}_j =0$), 
which we call a junction.
Here, we consider the simple case that 
the three types of the color-electric charge
are placed on the corners of a regular triangle.
The non-vanishing plaquettes $\hat{\Sigma}_{x,1\;\mu\nu}$,
$\hat{\Sigma}_{x,2\;\mu\nu}$ and $\hat{\Sigma}_{x,3\;\mu\nu}$ 
are properly included so as to minimize the length of the 
color-electric Dirac string, which corresponds to the energy 
minimization condition.
Then, the position of the junction is given by the Fermat 
point \cite{Kamizawa:1993hb}.
As a result, we get a typical $Y$-shaped flux-tube object in 
U(1)$^2$ DGL theory, {\it i.e.} the baryonic flux tube.

\par
\begin{figure}[t]
\centerline{\epsfxsize = 7cm\epsfbox{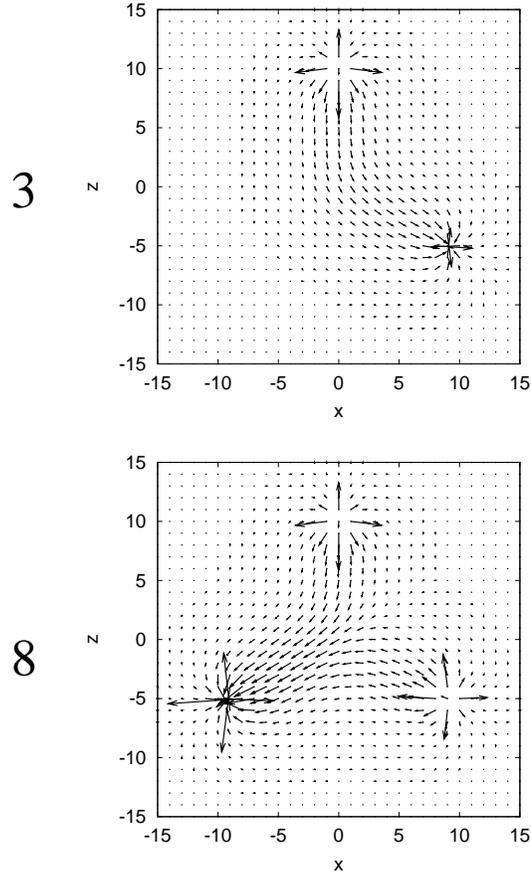}}
\caption{The profiles of the color-electric field in the Cartan
representation for 3- (upper) and 8- (lower) components
in the baryonic flux tube in the $x$-$z$ plane at $y=0$.
The junction and the quarks are located at $(x,y,z)=(0,0,0)$, and
$R(0,0,9)$, $B(9,0,-5)$, $G(-9,0,-5)$, respectively.}
\label{fig:38-baryon-ele}
\end{figure}
\begin{figure}[t]
\centerline{\epsfxsize = 7cm\epsfbox{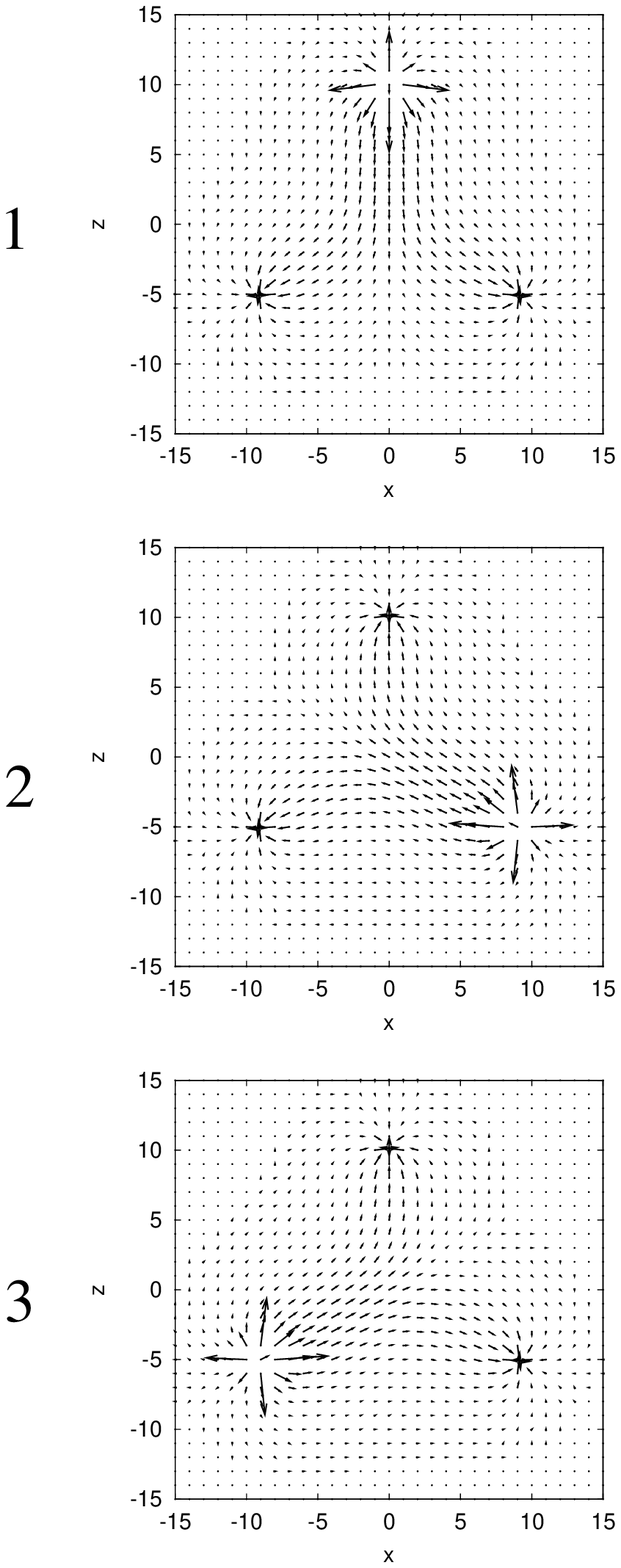}}
\caption{The profiles of the color-electric field in the color-electric 
representation, expressed on the weight
vectors of the SU(3) algebra, $\vec{w}_1$ (upper),
$\vec{w}_2$ (middle), and $\vec{w}_3$ (lower)
in the baryonic flux-tube system in the $x$-$z$ plane at $y=0$.
The junction and the quarks are located at $(x,y,z)=(0,0,0)$, and
$R(0,0,9)$, $B(9,0,-5)$, $G(-9,0,-5)$, respectively.}
\label{ele-baryon-ele}
\end{figure}
\begin{figure}[t]
\centerline{\epsfxsize = 7cm\epsfbox{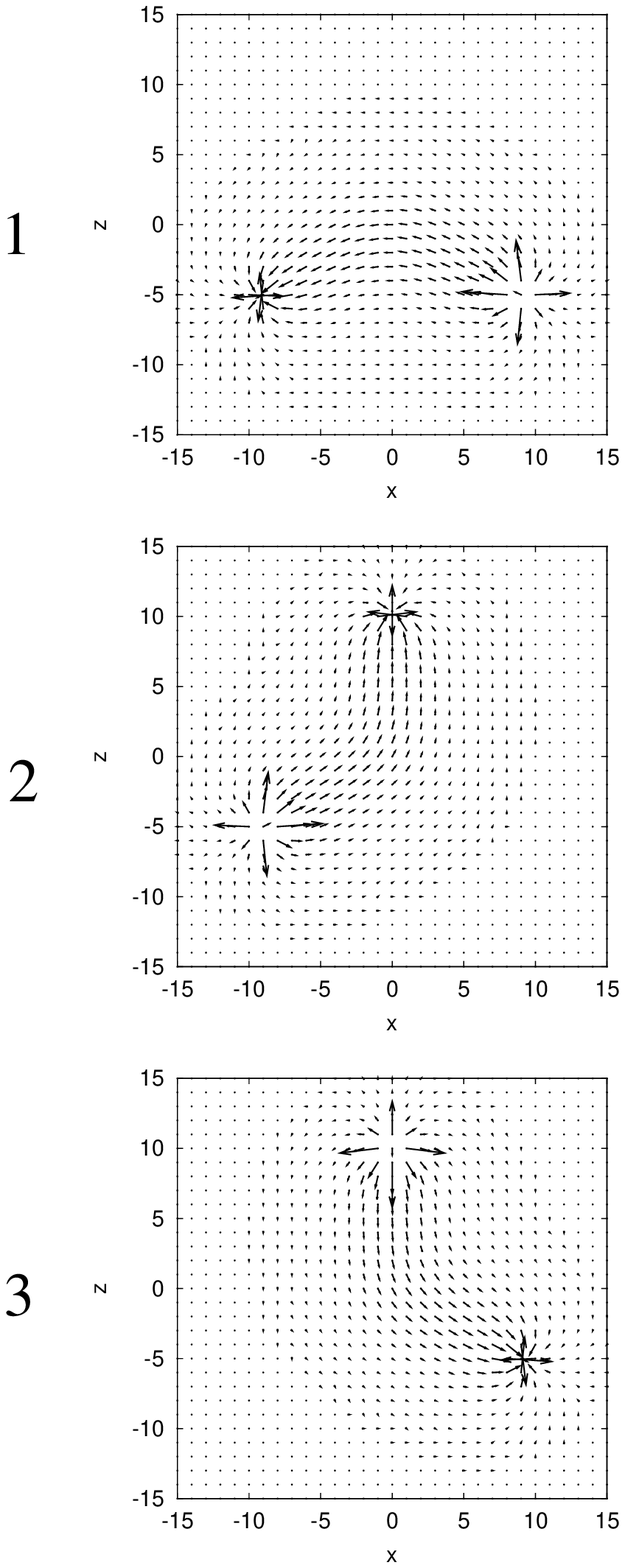}}
\caption{The profiles of the color-electric field in the color-magnetic
representation, expressed on the root
vectors of the SU(3) algebra, $\vec{\epsilon}_1$ (upper),
$\vec{\epsilon}_2$ (middle), and $\vec{\epsilon}_3$ (lower)
in the baryonic flux-tube system in the $x$-$z$ plane at $y=0$.
The junction and the quarks are located at $(x,y,z)=(0,0,0)$, and 
$R(0,0,9)$, $B(9,0,-5)$, $G(-9,0,-5)$, respectively.}
\label{fig:mag-baryon-ele}
\end{figure}
\begin{figure}[t]
\centerline{\epsfxsize = 7cm\epsfbox{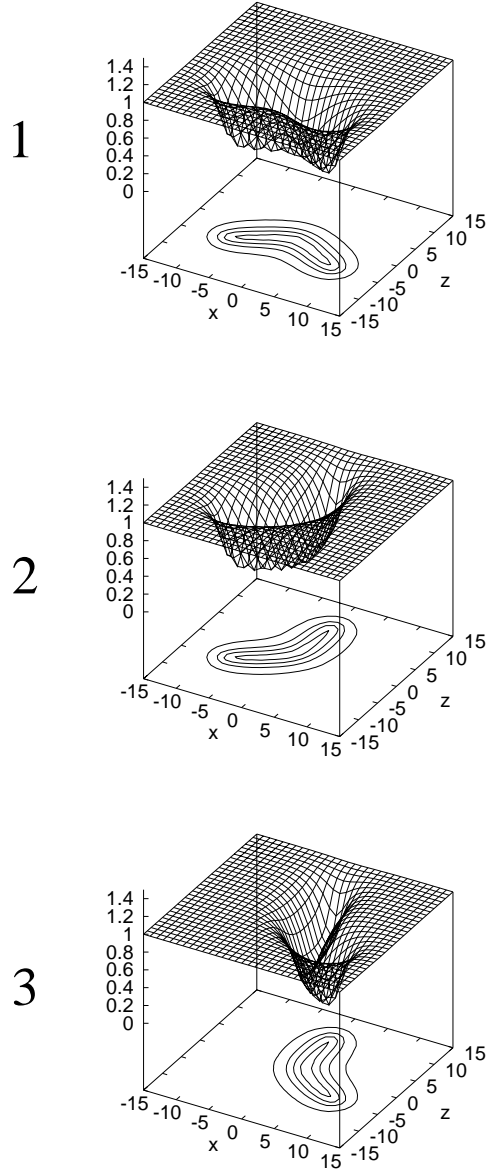}}
\caption{The profiles of the Higgs field of $|\chi_1|$ (upper),
$|\chi_2|$ (middle), and $|\chi_3|$ (lower)
in the baryonic flux-tube system in the $x$-$z$ plane at $y=0$.
The junction and the quarks are located at $(x,y,z)=(0,0,0)$, and
$R(0,0,9)$, $B(9,0,-5)$, $G(-9,0,-5)$, respectively.}
\label{fig:baryon-higgs}
\end{figure}
\begin{figure}[t]
\centerline{\epsfxsize = 10cm\epsfbox{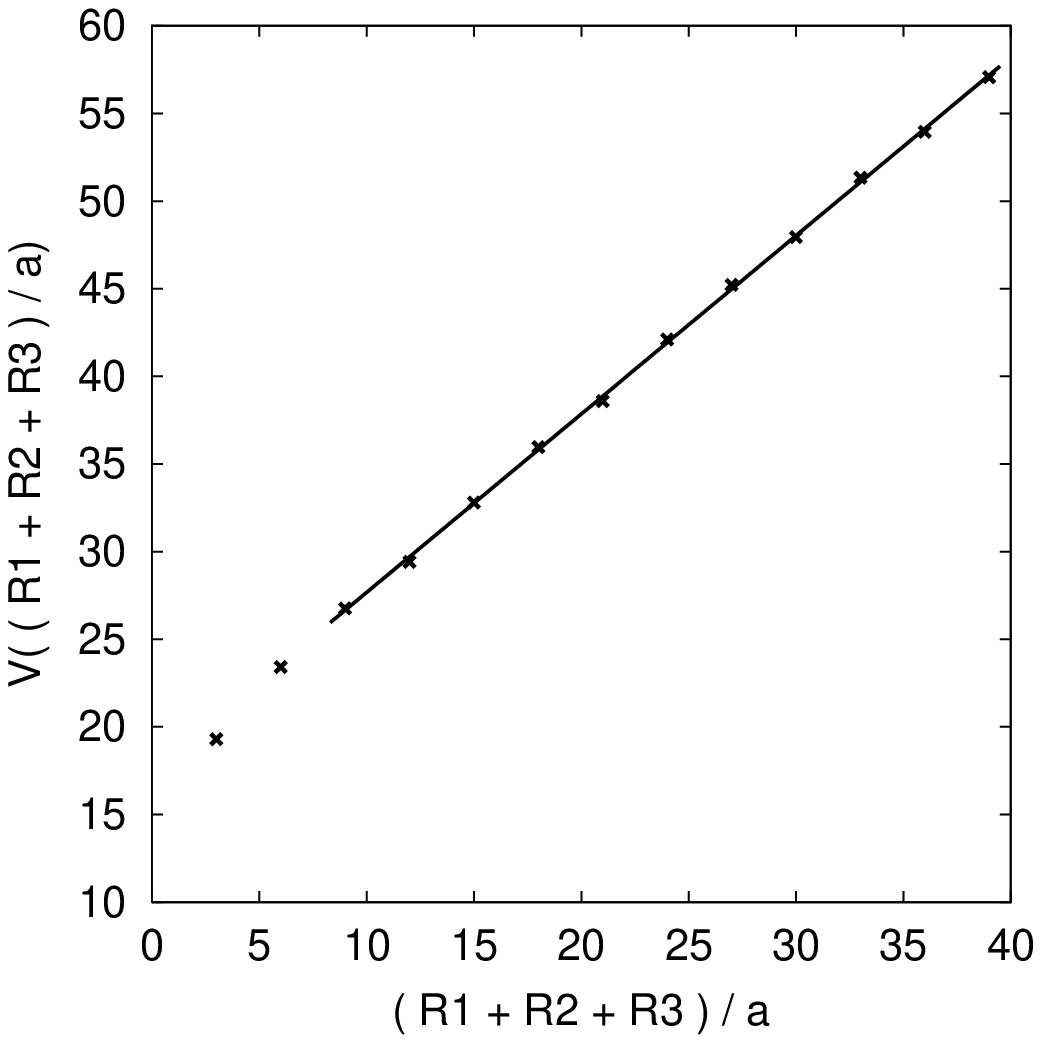}}
\caption{The three-quark potential in the [U(1)]$^2$ DGL theory, where 
$R_i=|\bvec{x}_i-\bvec{x}_J|$.
The parameter set is taken as 
$\beta = 1$, $\hat{m}_B = \hat{m}_\chi =0.5$.}
\label{fig:baryon-pot}
\end{figure}

In Figs.~\ref{fig:38-baryon-ele}-\ref{fig:mag-baryon-ele}, 
we show the profiles of the color-electric field
corresponding to the Cartan, the color-electric, and the color-magnetic 
representations of the dual gauge field.
The Weyl symmetric structure can be observed in the last two 
representations.
The monopole field does not depend on which representation is chosen, 
for the same reason as in the discussion of the mesonic flux tube, which
is shown in Fig.~\ref{fig:baryon-higgs}.
One finds that all of these profiles faithfully reflect the
structure of the color-electric Dirac string.
The potential is obtained analogously to the mesonic system, which 
is shown in Fig.~\ref{fig:baryon-pot}.
Here,  parametrizing the potential of the linear part as
\be
V(\bvec{x}_1,\bvec{x}_2,\bvec{x}_3) \sim 
\sigma^L \sum_{i=1}^3 | \bvec{x}_i - \bvec{x}_J |,
\ee
where $\bvec{x}_i$ and $\bvec{x}_J$ denote the position of the quarks and 
of the junction on the dual lattice, respectively, 
we can extract the string tension $\sigma^L$.
One finds that this is almost reproduced by the analytical one, 
since $\sigma^L \sim 1.0 \sim 4 \beta \pi\hat{m}_B^2/3$.
It is interesting to note that while  each profile of
the color-electric field in the color-electric 
representation has similar form to the 8-flux in the Cartan representation,
the color-electric field in the color-magnetic representation provides 
the 3-flux type structure. It is, of course, possible to
study the energy and the field distribution corresponding to 
different shapes of baryonic flux tube in a static configuration.

\section{Summary and discussion}

We have studied the classical flux-tube solutions for 
the mesonic and the baryonic states within the dual Ginzburg-Landau 
(DGL) theory by using the dual lattice formulation in
the Weyl symmetric approach.
The color-electric Dirac string singularity, which determines 
the filament core inside the flux tube, has been treated 
as a connected stack of quantized plaquettes which lead to the 
phase $\pm 2\pi$ in the dual lattice formulation.
This formulation is flexible to reproduce various shapes of the flux tube
just by putting the quantized plaquettes which are pierced by a  
color-electric Dirac string of any form.
We have found that the manifestly Weyl symmetric approach, 
in particular, the color-magnetic representation of the dual gauge field 
is the most convenient one to investigate flux-tube solutions 
in the [U(1)]$^2$ DGL theory, since 
this gives a quite similar form with the U(1) case \cite{Koma:2000wn}.

\par
In this paper we have concentrated on formulating a 
simple method to investigate the qualitative properties of
the classical flux-tube solution in the U(1) and [U(1)]$^2$ DGL theory.
This work can be extended to the study of the
flux tube in the quantized DGL theory by using the Monte Carlo method
in four dimensional Euclidean space time.
Then, also the effect of string fluctuations becomes a possible topic 
of investigation.
Even without string fluctuations, 
we can discuss more quantitative properties of the 
hadronic flux tubes, based on a quantum DGL theory. 
This is under preparation.
The application of this formulation to the flux-tube ring 
solution as the glueball state \cite{Koma:1999sm} is also interesting.

\par 
A crucial criterion for a viable confinement mechanism is the ability 
to reproduce the Casimir scaling of the forces at intermediate distances
between static charges in different representation,
$F_{R_i}/F_{R_j}=C_2(R_i)/C_2(R_j)$ with $C_2$ as the eigenvalue of the
quadratic Casimir operator $\lambda_j^2$ in the representations $R_i$ 
and $R_j$, respectively.  For the ratio of adjoint to fundamental charges 
in SU(3) gauge theory, this would 
give $\sigma_{\mathrm adj}/\sigma_{\mathrm fund}=9/4$.
Casimir scaling is in the discussion since the first lattice indications
for it appeared in the eighties~\cite{early_works}, and at that time it
was challenging for the bag model~\cite{hansson}.
Enhanced attention recently, due to the lively discussion of competing 
confinement mechanisms, in Ref.~\cite{Deldar:1999vi} the string tensions
of the fundamental and the adjoint representations were
computed and the ratio came out to be nearly 2, which was close to 9/4.
On the other hand, in Ref.~\cite{Bali} they have studied
the ratio of entire potentials including Coulomb and constant 
terms in addition to the linear term and the ratio turns out to be
very close to 9/4.
All detailed and microscopic mechanisms of confinement find it hard to 
explain this observation, while it is more natural from the point of view 
of the semi-phenomenological Stochastic Vacuum Model~\cite{SVM}. 

\par 
Discussing the Abelian projection in terms of $\lambda_3$ and $\lambda_8$
would suggest to evaluate $\sigma_{\mathrm adj}/\sigma_{\mathrm fund}$ 
with diagonal gluons only, giving a ratio 3. 
This makes it hard to understand why Casimir scaling should hold in  
Abelian projected gluodynamics.
Considering the DGL theory just at a phenomenological level, it would
be sufficient to restrict it exclusively to mesonic, baryonic, glueball 
and perhaps to exotic states, and it would be inappropriate to apply it 
to the so-called gluelump bound states made of infinitely heavy adjoint 
charges. However, because of the current interest, it might be amusing 
to consider briefly how this kind of string would be represented within 
the DGL theory.
Although this theory, as an effective theory of gluodynamics with external
charges, is constructed referring to Abelian projection, we are free to look 
at the Casimir problem afresh. 

\par
In fact, the DGL theory is rather promising to discuss 
the Casimir scaling problem
without extra effort. To see this, we recall that the DGL theory represents 
the mesonic string as degenerate $R-\bar{R}$, $B-\bar{B}$ and $G-\bar{G}$ 
colored states. 
In the same spirit it is natural to represent gluelump strings as stretching
out between pairs of adjoint charges, each of them being made out of quark and 
antiquark as $B\bar{G}-\bar{B}G$, $G\bar{R}-\bar{G}R$, and $R\bar{B}-\bar{R}B$ 
states. Thus, it is rather a string formed by two pairs with their respective 
Dirac strings superposed. In the Bogomol'nyi limit one directly gets the ratio 
$\sigma_{\mathrm adj}/\sigma_{\mathrm fund}$  
using the manifest Weyl invariant formulation of the DGL 
theory~\cite{Koma:2000wn}.
In this limiting case the ratio is equal to 2, reflecting the presence of 
two independent color-electric Dirac strings inside the adjoint flux tube.
Entering the type-II dual superconductor parameter range, the ratio will 
increase, while decreasing towards the type-I region. Our studies show that 
the ratio of string tensions depends only on the ratio between dual vector 
and monopole mass, via $\kappa=m_{\chi}/m_B$. It has been conjectured that 
the ratio 9/4 is reproduced in a certain type-II vacuum.

\section*{Acknowledgment}
Y.K. and H.T. are grateful to H. Suganuma for his useful comment on the 
Weyl symmetry in the DGL theory.
Y.K. thanks M. Takayama for useful discussions.
E.-M.I. acknowledges discussions with M. M\"uller-Preussker.
He is grateful for the support by the 
Ministry of Education, Culture and Science of Japan (Monbu-sho) 
providing the opportunity to work at RCNP.
T.S. acknowledges the financial support from JSPS Grant-in Aid
for Scientific Research (B) No. 10440073 and No. 11695029.

\begin{table}
\begin{center}
\begin{minipage}{12cm}
\caption{\small 
The color-electric Dirac string structure in the dual gauge field
in $q$-$\bar{q}$ system for various representations 
(for Figs.~\ref{fig:38-meson-ele}-\ref{fig:meson-higgs}) 
is schematically summarized. 
Here, $\uparrow$ and $\downarrow$ correspond to the one 
Dirac string singularity.
If we circulate around these singularities as counter-clockwise way, 
they lead to the phase $+2\pi$ and $-2\pi$, respectively.
Notice that $\Uparrow = 2 \times \uparrow$ and 
$\Downarrow = 2 \times \downarrow$.}
\vspace{0.3cm}
\end{minipage}
\begin{tabular}{ccc c  ccc c  ccc}
\hline
 & \multicolumn{2}{c}{3-8 basis}
 & & \multicolumn{3}{c}{electric basis} 
 & & \multicolumn{3}{c}{magnetic basis}
\\
\hline
   & $B_{\mu}^3$     & $B_{\mu}^8$
 & & $B_{1\;\mu}^e$ & $B_{2\;\mu}^e$  & $B_{3\;\mu}^e$
 & & $B_{1\;\mu}^m$ & $B_{2\;\mu}^m$ & $B_{3\;\mu}^m$
\\
\cline{2-3}
\cline{5-7}
\cline{9-11}
$R-\bar{R}$  & $\downarrow_c$  & $\downarrow_c$
& & $\Downarrow_e$ & $\uparrow_e$ & $\uparrow_e$
& & $0$  & $\uparrow_m$ & $\downarrow_m$ \\
$B-\bar{B}$  & $\uparrow_c$ & $\downarrow_c$ 
& &$\uparrow_e$ & $\Downarrow_e$ & $\uparrow_e$  
& &$\downarrow_m$ & $0$ & $\uparrow_m$\\
$G-\bar{G}$  & $0$            & $\Uparrow_c$
& &$\uparrow_e$ & $\uparrow_e$ & $\Downarrow_e$  
& & $\uparrow_m$ & $\downarrow_m$ & $0$\\
\hline
\end{tabular}
\end{center}
\end{table}


\end{document}